\documentclass[%
 reprint,
showpacs,
 amsmath,amssymb,
 aps,
 pra,
floatfix,
superscriptaddress
]{revtex4-1}
\usepackage[utf8]{inputenc}
\usepackage[T1]{fontenc}
\usepackage[]{graphicx}
\usepackage{graphicx}
\usepackage{color}
\usepackage[dvipsnames]{xcolor}
\usepackage[english]{babel}
\usepackage[normalem]{ulem}
\usepackage[bookmarks,bookmarksopen,bookmarksdepth=2]{hyperref}
\hypersetup{
    colorlinks=true,
    citecolor=blue,
    linkcolor=blue,
    filecolor=magenta,
    urlcolor=cyan,
}

\usepackage{microtype}
\usepackage{braket}
\usepackage{esvect}
\usepackage[export]{adjustbox}
\usepackage{dsfont}
\usepackage[T1]{fontenc}
\usepackage[utf8]{inputenc}
\usepackage{url}
\usepackage{orcidlink}

\usepackage{amsmath,amsfonts,amssymb,amsthm}
\usepackage{mathtools}
\usepackage{bbm}
\usepackage{mathrsfs,physics,xfrac,cancel,tensor,relsize,tikz,lipsum,quantikz}

\newcommand{\col}[2]{\|#1\rangle\langle#2\|}

\begin{document}

\title{Using an SU(3)/U(2) Wigner Function to Represent Noisy Spin Ensembles}
\author{Andrew Kolmer Forbes\orcidlink{0009-0003-8730-8007}}
\email{aforbes@unm.edu}
\affiliation{Center for Quantum Information and Control, Department of Physics and Astronomy, University of New Mexico, Albuquerque, New Mexico 87131, USA}

\begin{abstract}

The SU(2) Wigner function represents a quantum state of a spin-$J$ as a real-valued function on the surface of a 2-sphere. For an ensemble of $N$ spin-1/2 particles, this representation is useful when the dynamics is restricted to a single SU(2) irrep, e.g., the symmetric subspace with $J=N/2$. Physically relevant noise sources tend to be local, such as spontaneous emission, depolarizing, and incoherent optical pumping, all of which transfer the state outside of the initial irrep, and as such the SU(2) Wigner function is no longer a useful representation. In this work, we address this issue by encoding a noisy spin ensemble in an SU(3) irrep, and evaluating the SU(3) Wigner function for that irrep. We find that physical constraints enforced by the noise eliminate all but three real parameters from the input to the Wigner function, which can then be interpreted as a polar, azimuthal, and radial component. This interpretation leads us to refer to the resulting Wigner function as the \emph{solid spin Wigner function}, visualized on a solid ball rather than a hollow sphere.
    
\end{abstract}

\maketitle

\section{Introduction}
In the Liouville formulation of classical mechanics, the state is represented by joint probability distributions over canonical coordinates in phase space. These distributions possess many desirable qualities, such as normalization, positivity, and marginalization. Such properties arise from the commutativity of classical observables, such as position and momentum. It is no surprise, then, that constructing similar distributions for quantum states leads one to realize they must relinquish some of these properties \cite{Moyal1949}.

Formulated by Hermann Weyl in the 1920s, the \emph{Weyl transformation} provides a bridge between operators in Hilbert space and classical functions in phase space \cite{Weyl1918,Weyl1927}. Soon after, Eugene Wigner developed the now-famous Wigner function \cite{Wigner1932}, which utilized the Weyl transformation to map a bosonic state to a real-valued distribution of position and momentum. While real-valued, this function possessed the property of admitting negative values, something classical joint probability distributions do not exhibit. Nonetheless, the distribution is both normalized and can be marginalized to reconstruct probability distributions of position and momentum. Thus, the fitting name of \emph{quasiprobability distributions} is assigned to such functions. There exists quasiprobability distributions \cite{Glauber1963,Husimi1940}, such as the Husimi-Q distribution, which regain positivity, but which come at other costs. Despite the work on bosonic Wigner functions in flat phase space, a Weyl map to spaces representing arbitrary Lie groups (which may have curved geometry) was not established until 1957 by Ruslan Stratonovich \cite{Stratonovich1957}. This work laid the foundation for the SU(2) Wigner function, first defined in 1981 by Agarwal \cite{Agarwal1981}.

Far from a pure mathematical quirk, these negative regions of the Wigner function convey information about the quantum nature of the state \cite{Hudson1974,Hillery1984,Salazar2023}. Wigner negativity is related to the noncontextual nature of quantum mechanics \cite{Spekkens2008}, and in recent years it has been shown to be a resource for the generation of nonclassical bosonic states \cite{Albarelli2018}, and is necessary for universal bosonic quantum computation \cite{Mari2012,Veitch2012,Delfosse2015}. Unsurprisingly, noise tends to diminish this negativity, corresponding to quantum states decohering into ones which look more ``classical''.

In bosons and spins of size-$J$ (total angular momentum $J$), Markovian noise can typically be described tractably, as it leaves the dimension of the Hilbert space unchanged. For example, in a spin-$J$ system Markovian noise can be described using jump operators $\hat J_\pm$ which preserve $J$. However, many large spin-$J$ states are not truly a single spin, but rather the symmetric subspace of $N=2J$ spin-1/2 particles \cite{Dicke1954}. This leads to a problem when considering the spin Wigner function, as the physically relevant noise in these systems is usually local (i.e., incoherent optical pumping, spontaneous emission, depolarizing, etc.), which breaks the $J$-symmetry of the state. While the Hilbert space of the symmetric subspace scales linearly in $N$, the dimension of the space needed to describe a system undergoing arbitrary local noise scales exponentially, formed by the tensor product of all spin-1/2 particles. Thus, the usual spin Wigner function for a spin-$J$ system cannot be used to describe the state once local noise occurs.

The na\"ive construction of the Wigner function for this exponentially large space is the product of spin Wigner functions for each spin-1/2 particle, resulting in a $2N$-dimensional parameter space. While satisfying the basic properties of a Wigner function, this construction is typically intractable for moderate $N$, and lacks the convenient visualizations available to the spin-$J$ Wigner function. Motivated by this, we aim to develop a Wigner function which applies to spin-$J$ systems undergoing local noise while avoiding the na\"ive product construction.

A challenge of defining this new Wigner function is that the system ($N$ spin-1/2 particles) naturally obeys SU(2) symmetry, but the Hilbert space is larger than a single SU(2) irrep. Our approach to solving this problem is to embed the state in an irreducible representation of a larger group, which contains as a subgroup SU(2). By doing so, we increase the size of the representation space, while preserving the SU(2) symmetry of the underlying system being represented. We will demonstrate that SU(3) is a suitable contender for this larger group, as it contains three SU(2) subgroups, and permits an irreducible representation which is isomorphic to the space used to describe the noisy SU(2) system.

In this work we will assume symmetry of local noise, which restricts the dynamics to states which obey permutation symmetry. Such dynamics can be described using the collective state basis \cite{Chase2008}, which scales quadratically in $N$ \cite{Zhang2018}. This basis contains no degeneracy, and we show that it is isomorphic to a direct sum of spin-$J$ Hilbert spaces. This isomorphism allows for noisy states in the collective state basis to be embedded in an irreducible representation (irrep) of SU(3), which preserves the natural group structure of the noisy state by making use of SU(2) subgroups within SU(3). Through this embedding, one can describe an SU(3) Wigner function \cite{Tilma2011,Klimov_SUn} with special properties arising from physical constraints on the collective state basis. These properties allow for simplifications in the number of parameters, reducing the total number from four real parameters to three. These three parameters take the role of the usual polar and azimuthal angles, with the third parameter having an interpretation as a radial component. This leads us to think of the resulting Wigner function as a \emph{solid} spin Wigner function, which exists not on the surface of a sphere but on a ball of unit radius.

\section{Formalism}
\label{sec:formalism}

In this section we will introduce the formalism needed to understand the collective state basis used for modeling local noise, the SU(3) group and its irreducible representations, and Moyal quantization, each of which is necessary for constructing the solid spin Wigner function. In Sec.~\ref{sec:moyal_quantization} we will review the concept of Wigner functions and phase space formalism from the generalized perspective of Moyal quantization and Lie groups. Then, in Sec.~\ref{sec:collective_states} we will review the collective state basis for SU(2) systems undergoing symmetric local noise. In Sec.~\ref{sec:contructing_v_main_text} we will show how this basis is isomorphic to a Hilbert space constructed from a nondegenerate direct sum of SU(2) irreps, and finally in Sec.~\ref{sec:su3} we will introduce the algebra of $\mathfrak{su}(3)$, and define its corresponding group structure and representations.

\subsection{Moyal Quantization}
\label{sec:moyal_quantization}

The phase space formulation of quantum mechanics, referred to in many modern descriptions as Moyal quantization, maps an operator $\hat f$ in Hilbert space $\mathcal H$ to a symbol $W_{\hat f}(\Omega)$, where $\Omega$ is a point in phase space. This mapping is generated by computing the trace of $\hat f$ with a kernel $\hat\omega(\Omega)$,
\begin{equation}
    W_{\hat f}(\Omega)=\Tr(\hat\omega(\Omega)\hat f).
\end{equation}
If $\hat f$ is a density operator then $W_{\hat f}(\Omega)$ is referred to as a \textit{Wigner function}, and $\hat\omega(\Omega)$ must obey the so-called Stratonovich-Weyl (SW) conditions which enforce several physically reasonable restrictions on Moyal quantization, first introduced by Stratonovich \cite{Stratonovich1957}. One such formulation of these conditions is \cite{Biff_and_Mann}
\begin{enumerate}
    \item Linearity: $\hat f\mapsto W_{\hat f}(\Omega)$ is linear and bijective,
    \item Hermiticity: $\hat\omega(\Omega)=\hat\omega(\Omega)^\dagger$,
    \item Standardization: $\Tr(\hat\omega(\Omega))$=1,
    \item Covariance: $\hat U(g)\hat\omega(\Omega)\hat U(g)^{-1}=\hat\omega(g\cdot\Omega)$,
    \item Traciality: $\int_MW_A(\Omega)W_B(\Omega)\mathrm d\Omega=\Tr(\hat A\hat B)$,
\end{enumerate}
where $g\in G$ is an element of some group $G$ and $\mathrm d\Omega$ is the $G$-invariant measure on a manifold $M$. Hermiticity and standardization enforce that the Wigner function is both real-valued and normalized. The covariance condition enforces that the action of $g$ acting on a point $\Omega$ in the manifold is the same as the corresponding unitary in Hilbert space $\hat U(g)$ acting on the kernel. Finally, the traciality condition for operators $\hat A$ and $\hat B$ in $\mathcal H$ allows one to recover expectation values from the phase space representation. When these operators are not density operators we will refer to their Moyal quantized symbol as their \textit{Weyl symbol}. Kernels satisfying the SW conditions can be constructed when an underlying Lie algebra of a Hilbert space can be identified \cite{Biff_and_Mann}. Further, explicit construction of Wigner functions for arbitrary SU($n$) Hilbert spaces \cite{Klimov_SUn} has been achieved. 

The manifold $M$ which $\Omega$ is a point on, can be found by identifying the underlying group structure of unitaries on the Hilbert space. In general, if a Hilbert space $\mathcal H_g$ forms a representation of a Lie algebra $\mathfrak{g}$, we can identify a Lie group $G$ with elements $g$. We can further define a fiducial state $\ket{\psi_0}\in\mathcal H_g$ such that all group actions on $\ket{\psi_0}$ can be broken into two pieces $g=B(\Omega) h$. Here, $h\in H$ is an element of the isotropy subgroup, where $h\ket{\psi_0}=e^{-i\phi_h}\ket{\psi_0}$. The group element $B(\Omega)\in G/H$ is an element of a coset space, and uniquely identifies each point in $G/H$ with a point $\Omega$ on a manifold. By allowing $B(\Omega)$ to act on the fiducial state $\ket{\psi_0}$, we can associate each point on the manifold with a ``coherent state'' $B(\Omega)\ket{\psi_0}=\ket{\Omega}$ in $\mathcal H_g$~\cite{Biff_and_Mann}.

For the Wigner function of a quantum harmonic oscillator, points in phase space associated with the Weyl-Heisenberg group are given by the complex amplitude $\alpha=(x+ip)/\sqrt{2}$ [for canonical coordinates $(x,p)$], which define the mean values of coherent states $\ket{\alpha}$. These are obtained by applying the coset action of the displacement operator $\hat D(\alpha)$ to the fiducial state, the vacuum state $\ket0$. Similarly, for systems with SU(2) symmetry, points in phase space are defined using the polar and azimuthal angles $(\theta,\phi)$ which correspond to the rotated spin coherent states (SCS) $\ket{\theta,\phi}$. These states are obtained by applying the coset action $B(\Omega)\in$ SU(2)/U(1) of two rotation operators $\hat R_z(\phi)\hat R_y(\theta)$ about $y$ and $z$ sequentially to the highest weight state of an SU(2) irreducible representation. The choice of fiducial state is important when considering the construction of Wigner functions, as it can change the coset space used to define the manifold. From the choice of a fiducial coherent state, along with the identification of a Lie algebra and its corresponding group structure manifold, a Wigner function can be constructed which maps the state in the representation of the algebra to the continuous manifold of the group.

\subsection{The Collective State Basis}
\label{sec:collective_states}

For a collection of $N$ spin-1/2 particles, the total Hilbert space $\mathcal H_N$ has dimension $\text{dim}(\mathcal H_N)=2^N$. This space is the tensor product space of $N$ 2-level Hilbert spaces $\mathcal H_{1/2}$, and can be decomposed into a direct sum of SU(2) irreps $\mathcal H_{J,\alpha}$ as
\begin{equation}
    \bigotimes_{i=1}^N\mathcal H_{1/2}=\bigoplus_{J=\{0,1/2\}}^{N/2}\bigoplus_{\alpha=1}^{d_J^N}\mathcal H_{J,\alpha},\label{eq:irrep_decomp}
\end{equation}
where $J$ begins at either 0 or 1/2 depending on the parity of $N$, and $\alpha$ is a degeneracy label running from 1 up to the degeneracy $d_J^N$ given by \cite{Chase2008}
\begin{equation}
    d_J^N=\frac{2J+1}{N/2+J+1}\binom{N}{N/2-J}.
\end{equation}
The size of each irrep is $\text{dim}(\mathcal H_{J,\alpha})=2J+1$. The largest irrep, which has $J=N/2$, contains no degeneracy and is known as the \emph{symmetric subspace} $\mathcal H_{\text{sym.}}$. This subspace is the space of permutationally symmetric pure states $\mathcal H_{\text{sym.}}\subset \mathcal H_N$ and has dimension scaling only linearly with $N$, $\text{dim}(\mathcal H_{\text{sym}})=N+1$. Unlike the total Hilbert space, states in the symmetric subspace can be labeled uniquely by the nondegenerate eigenvalues of the projective angular momentum operator $\hat J_z$ and the total angular momentum $\hat J^2=\hat J_x^2+\hat J_y^2+\hat J_z^2$, where
\begin{equation}
    \hat J_\nu=\frac12\sum_i\hat\sigma_\nu^{(i)}\label{eq:collective_operator}
\end{equation}
and $\nu\in\{x,y,z\}$ labels Pauli matrices $\hat\sigma_\nu^{(i)}$ on the $i$th particle. The basis labeled by these eigenvalues is the angular momentum eigenstate basis, $\ket{J,M}$, where
\begin{align}
    \hat J^2\ket{J,M}&=J(J+1)\ket{J,M},\\
    \hat  J_z\ket{J,M}&=M\ket{J,M}.
\end{align}
In the symmetric subspace $J=N/2$, but outside of this subspace where $J<N/2$ these states are degenerate, and require a third label $\alpha$ to label uniquely. The degeneracy label $\alpha$ is not physical in the sense that it does not correspond directly to an observable of the system. Rather, $\alpha$ is a label which counts the number of unique ways to arrange $N$ spins such that their total angular momentum is $J$, and their projective angular momentum is $M$. Thus, $\alpha$ is fundamentally a label corresponding to the local structure of a state, which can only be made meaningful if a particular tensor product construction of $\ket{J,M,\alpha}$ is defined \textit{a priori}.

\begin{figure}
    \centering
    \includegraphics[width=0.94\linewidth]{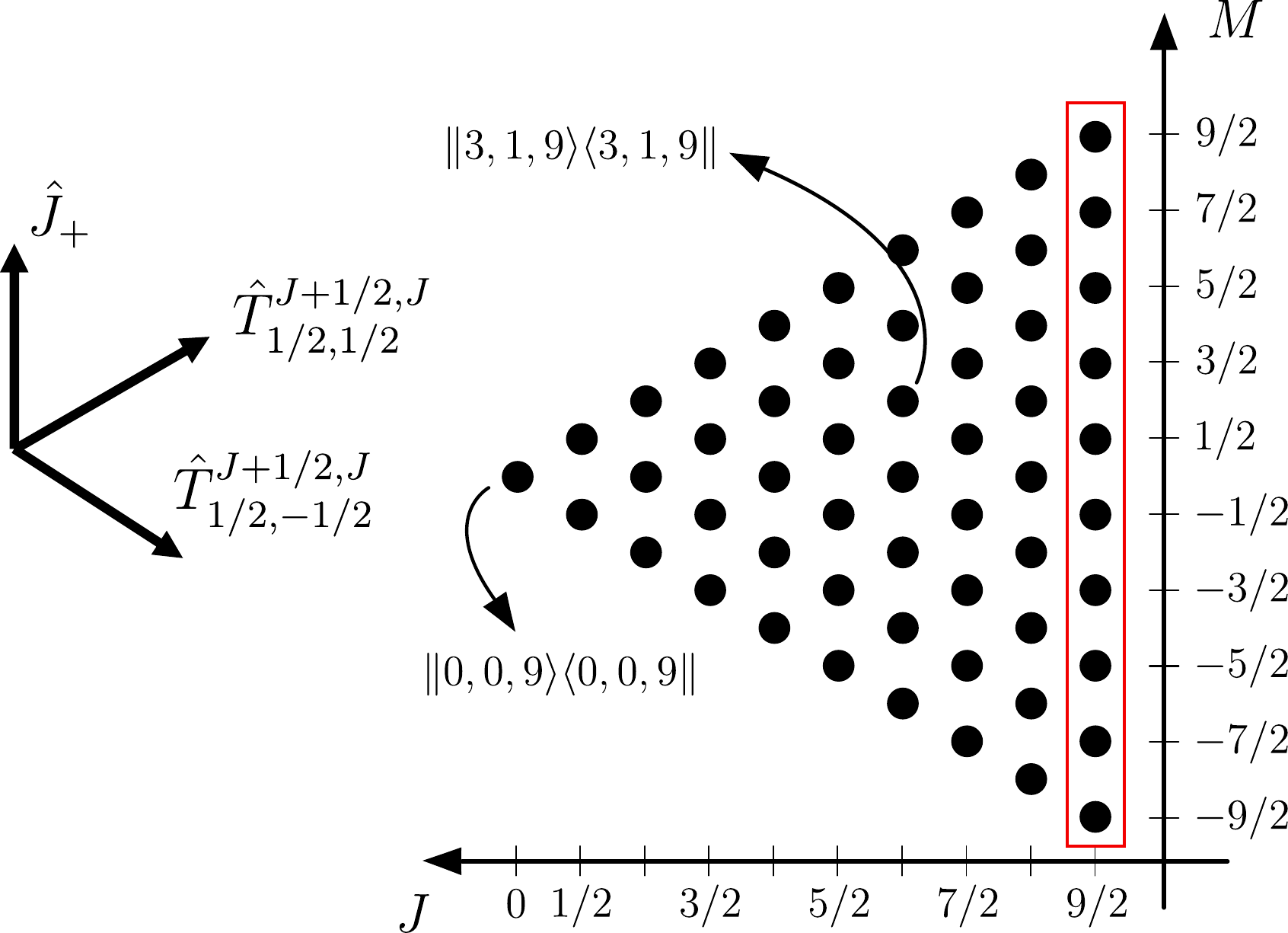}
    \caption{Diagram of the collective state basis, with half-integer SU(2) irreps included. The symmetric subspace, which is pure and nondegenerate, is indicated using a red circle. Two examples of collective states [Eq.~(\ref{eq:coll_state})] are labeled. Also depicted are the directions of raising operators $\hat J_+, \hat T_{1/2,1/2}^{J+1/2,J}$ and $\hat T_{1/2,-1/2}^{J+1/2,J}$ in this diagram, which are given by the spherical tensor operators (more precisely, given by weighted sums across $J$ of spherical tensor operators). These operators and their hermitian conjugates arise naturally as the jump operators for atom loss and gain channels \cite{Forbes2024,Zhang2018}.}
    \label{fig:coll_space}
\end{figure}

A basis for all permutationally symmetric operators on the total Hilbert space can be constructed using the \textit{collective state basis}, which is spanned by the basis elements
\begin{equation}
    \|J,M,N\rangle\langle J,M',N\|\equiv\frac{1}{d_J^N}\sum_{\alpha=1}^{d_J^N}\ketbra{J,M,\alpha}{J,M',\alpha}.\label{eq:coll_state}
\end{equation}
This is an operator basis which spans a subspace $\mathcal{C}$ of the space of linear operators on $\mathcal H_N$, $\mathcal B(\mathcal H_N)$. In particular, the space $\mathcal C$, spanned by Eq.~(\ref{eq:coll_state}), is the space of all permutationally symmetric linear operators on $\mathcal H_N$. The sum over $\alpha$ removes local information about the state, leaving the basis elements defined by Eq.~(\ref{eq:coll_state}) invariant under exchange of particles. For this reason, states undergoing symmetric local Markovian noise can be expanded in the collective state basis. A diagram of the collective state basis for $N=9$ qubits is pictured in Fig.~\ref{fig:coll_space}. A few example states have been labeled for clarity. The symmetric subspace, where $J=N/2$, has no degeneracies and is indicated using a red box. Note that there are states in Fig.~\ref{fig:coll_space} which differ in $J$ and $M$ by half-integer amounts. This is not typically physical, as most local noise channels can only increase or decrease angular momentum in integer amounts. However, atom loss and gain can lead to channels which can be represented by the half-integer structure pictured here \cite{Zhang2018}. Further, we will find later that including these half-integer spaces is convenient for the mapping to a representation of SU(3).

\subsection{Operators on Collective State Space}
\label{sec:contructing_v_main_text}


Because the collective state basis spans the space of permutationally symmetric operators on $\mathcal H_N$, there does not exist a basis element which couples states between two irreps $J$ and $J'$. In particular,
\begin{align}
    \nonumber\big(\col{J',M',N}{J,M,N}\big)&\col{J,M,N}{J,M,N}\\
    &\neq\col{J',M',N}{J,M,N}.\label{eq:j_diad}
\end{align}
However, superoperators $\Phi_{q,q'}^{J',J}$ do exist which transfer support between elements of different $J$ in the collective state basis, i.e.
\begin{align}
    \nonumber\Phi_{q,q'}^{J',J}&[\col{J,M,N}{J,M',N}]\\&=\col{J',M+q,N}{J',M'+q',N}.\label{eq:collective_state_map}
\end{align}
While operators like those seen in Eq.~(\ref{eq:j_diad}) do not exist in $\mathcal H_N$, it is apparent from previous works \cite{Forbes2024,Zhang2018} that linear operators which perform the action of Eq.~(\ref{eq:j_diad}) are useful, and can be practically implemented in simulations involving local noise. We will show in this section that one can define a mapping from operators in $\mathcal C$ (which is a vector space of operators) to operators which act on a new vector space $\mathcal V$. In this new space, the image of Eq.~(\ref{eq:coll_state}) can be interpreted as a genuine outer product. Both dynamics and expectation values are preserved across this mapping, and we find that linear operators which transfer population between irreps like Eq.~(\ref{eq:j_diad}) \emph{do exist} in this new vector space.

The space $\mathcal V$, which we refer to as the \emph{vectorized collective state basis} is defined as a direct sum of SU(2) irreps $\mathcal H_J$ of increasing total angular momentum $J$
\begin{equation}
    \mathcal V=\bigoplus_{J=0}^{N/2}\mathcal H_J,\label{eq:vectorized_collective_state_space}
\end{equation}
where the sum increases in half-integer amounts. The vector space $\mathcal V$ is thus spanned by vectors $\ket{J,M,N}\in\mathcal H_J$, and one can map elements in $\mathcal C$ to elements of $\mathcal B(\mathcal V)$, the space of linear operators on $\mathcal V$. This mapping promotes mixed states $\col{J,M,N}{J,M,N}\in\mathcal C$ to pure states $\ketbra{J,M,N}\in\mathcal B(\mathcal V)$.

The collective state basis, while spanning $\mathcal C$, is not self-dual, since the trace of a collective state acting on its hermitian conjugate is not a Kronecker delta function. Thus, the mapping from $\mathcal C$ to $\mathcal B(\mathcal V)$ on density operators $\hat\rho\mapsto\hat\rho_{\mathcal V}$ and linear operators $\hat A\mapsto\hat A_{\mathcal V}$ is made using both the collective state basis and its dual basis, given by
\begin{equation}
    \hat E_{M,M'}^{(J,N)}=d_J^N\col{J,M,N}{J,M',N},
\end{equation}
such that
\begin{subequations}
\begin{align}
    \hat\rho_{\mathcal V}&=\sum_{J,M,M'}\Tr(\hat\rho \left(\hat E_{M,M'}^{(J,N)}\right)^\dagger)\ketbra{J,M,N}{J,M',N}\label{eq:vectorized_rho}\\
    \nonumber\hat A_{\mathcal{V}}&=\sum_{J,M,M'}\Tr(\hat A\col{J,M',N}{J,M,N})\\
    &\;\;\;\;\;\;\;\;\;\;\;\;\;\;\times\ketbra{J,M,N}{J,M',N}.\label{eq:vectorized_A}
\end{align}
\end{subequations}
where $\ketbra{J,M,N}{J,M',N}\in\mathcal B(\mathcal V)$. Equations (\ref{eq:vectorized_rho}) and (\ref{eq:vectorized_A}) ensure that
\begin{equation}
    \Tr(\hat\rho \hat A)=\Tr(\hat\rho_{\mathcal V}\hat A_{\mathcal V})
\end{equation}
as desired. For a more detailed explanation of the appearance of the dual basis in this mapping, and why this particular mapping is natural, see App. \ref{app:ops_on_col}. In addition to preserving these properties of states and linear operators, the map $\mathcal C\mapsto\mathcal B(\mathcal V)$ also provides a representation of more general maps in $\mathcal C$ as linear operators on $\mathcal V$. Explicitly, one can write the map from Eq.~(\ref{eq:collective_state_map}) in $\mathcal B(\mathcal V)$ as
\begin{align}
    \nonumber\Phi_{q,q'}^{J',J}&[\col{J,M,N}{J,M',N}]\\
    &\;\;\;\;\;\mapsto \hat K_{q}^{J',J}\ketbra{J,M,N}(\hat K_{q'}^{J',J})^\dagger,
\end{align}
where $\hat K_{q}^{J',J}=\ketbra{J',M+q,N}{J,M,N}$. For the remainder of this work we will work in $\mathcal V$ and $\mathcal B(\mathcal V)$ to represent states undergoing local noise, and thus we will not use the double-bar notation introduced in Eq.~(\ref{eq:coll_state}) used to describe operators in $\mathcal C$.

It was shown in ref. \cite{Forbes2024} that maps corresponding to local noise jumps in $\mathcal C$ are proportional to generalized spherical tensor operators when mapped into $\mathcal B(\mathcal V)$ \footnote{The vector space $\mathcal V$ was not defined formally in \cite{Forbes2024}. However, operators which couple different $J$ were considered from an operational point of view, and are equivalent to the mapping to $\mathcal B(\mathcal V)$ presented here.}
\begin{equation}
    \hat T_{k,q}^{J',J}=\sqrt{\frac{2k+1}{2J'+1}}\sum_JC_{J,M,k,q}^{J',M+q}\ketbra{J',M+q,N}{J,M,N},\label{eq:sph_tensor}
\end{equation}
which satisfy the usual rotation property of spherical tensor operators \cite{Klimov2008}
\begin{equation}
\hat{R}_\nu(\theta) \, \hat{T}_{k,q}^{J,J'}\, \hat{R}_\nu^\dagger (\theta)= \sum_{q'} \hat{T}_{k,q'}^{J',J} D_{q',q}^{k,\nu}(\theta), \label{eq:GenSphericalRotProperty}
\end{equation}
where $\hat{R}_\nu(\theta)=e^{-i\theta\hat{J}_\nu}$ is the SU(2) rotation operator around the $\nu$-axis by the angle $\theta$, and $\hat D^{k,\nu}$ is the corresponding Wigner-$D$ matrix of rank $k$~\cite{Sakurai2011}. 

While the lack of an operator in $\mathcal C$ which couples collective states of different $J$ presented a challenge to defining jump operators within $\mathcal C$ (and led us to define $\mathcal V$ for convenience) we will soon see that this is an important property that can be used to simplify the SU(3) Wigner function.

\subsection{The SU(3) Group}
\label{sec:su3}
Unlike the SU(2) group, which is 3-dimensional with one Cartan operator, SU(3) is 8-dimensional with two Cartan operators. It is a simply connected and compact Lie group. The generators $\hat G_i$ of the Lie algebra $\mathfrak{su}(3)$ associated with SU(3) are proportional to the Gell-Mann matrices $\hat\lambda_i$, which are the 3-dimensional analog of the Pauli matrices \cite{GellMann1962}. Explicitly, $\hat G_i=\hat\lambda_i/2$, which satisfy the commutation relations,
\begin{equation}
    [\hat G_i,\hat G_j]=i\sum_{k}f_{ijk}\hat G_k,
\end{equation}
where $f_{ijk}$ is the structure constant of $\mathfrak{su}(3)$. 

Exponentiating $\mathfrak{su}(3)$ yields the corresponding Lie group SU(3), which has three SU(2) subgroups, with raising and lowering operators (ladder operators) in each that can be constructed from the generators $\hat G_i$. A representation of $\mathfrak{su}(3)$ can be spanned using these six ladder operators $\hat S_{ij}$ (with $i\neq j$ and $i,j\in\{1,2,3\}$) and two Cartan operators $\hat S_z^{\{12\}}$ and $\hat S_z^{\{23\}}$ which we will define soon. The $ij$ index on $\hat S_{ij}$ refers to the dyad which represents the operator in the 3$\times$3 fundamental representation of $\mathfrak{su}(3)$, i.e.,
\begin{equation}
    \hat S_{12}=\ketbra{1}{2}=\begin{pmatrix}
        0&1&0\\
        0&0&0\\
        0&0&0
    \end{pmatrix}.
\end{equation}
Thus, each non-ordered pair $\{ij\}$ for $i\neq j$ refers to an $\mathfrak{su}(2)$ subalgebra, with the raising operator $\hat S_{ij}$ when $j>i$, and the lowering operator when $j<i$. 

Using these ladder operators we can construct the equivalents of the $x$, $y$, and $z$ operators in each $\mathfrak{su}(2)$ subalgebra as \cite{Rowe1999}
\begin{subequations}
\begin{align}
    \hat S_x^{\{ij\}}&=\frac12\left(\hat S_{ij}+\hat S_{ji}\right)\\
    \hat S_y^{\{ij\}}&=\frac1{2i}\left(\hat S_{ij}-\hat S_{ji}\right)\\
    \hat S_z^{\{ij\}}&=\frac12\left[\hat S_{ij},\hat S_{ji}\right].
\end{align}\label{eq:plus_minus}
\end{subequations}
for $j>i$. We can choose our two Cartan operators $\hat h_1$ and $\hat h_2$ to be proportional to the $z$ operators of $\mathfrak{su}(2)_{\{ij\}}$ from any two SU(2) subgroups. In this work we will choose operators from the SU(2) subgroups $\{12\}$ and $\{23\}$ to span our Cartan subalgebra.

\begin{figure*}
    \centering
    \includegraphics[width=1\linewidth]{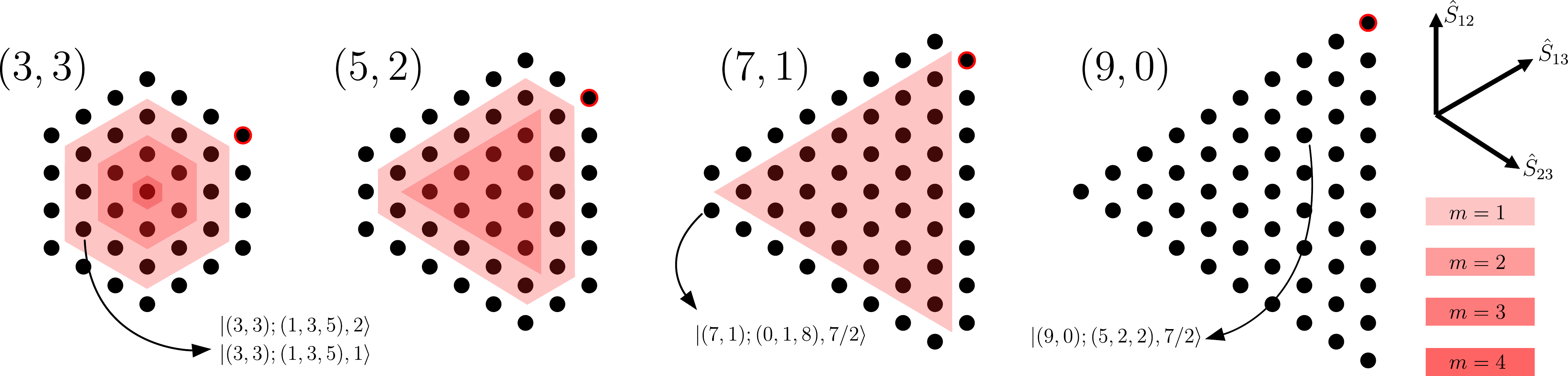}
    \caption{Weight diagrams of SU(3) irreducible representations labeled by Dynkin indices $(\lambda,\mu)$. Each dot represents a state in the representation, labeled by the occupation number $\boldsymbol{\nu}=(\nu_1,\nu_2,\nu_3)$. States in darker red regions have more multiplicity. The directions of the raising operators are shown on the right side of the figure. A few specific states $\ket{(\lambda,\mu);\boldsymbol{\nu}I}$ have been labeled for clarity. The highest weight state of each irrep is indicated with a red circle. Note that two states correspond to the weight labeled in the $(3,3)$ irrep, due to that weight having multiplicity 2. The multiplicity is lifted by the multiplicity label $I$, which is 2 for one of the states, and 1 for the other.
    }
    \label{fig:su3_irreps}
\end{figure*}

In order to label the weight states of $\mathfrak{su}(3)$ it is common to use eigenvalues of three operators belonging to the lifted Lie algebra $\mathfrak{gl}(3)$, which is the algebra of 3$\times$3 complex matrices. The operators commonly chosen are the diagonal operators $\hat S_{11}$, $\hat S_{22}$, and $\hat S_{33}$, with eigenvalues $(\nu_1,\nu_2,\nu_3)$. These are known as the \textit{occupation numbers} for the weight states of $\mathfrak{su}(3)\subset\mathfrak{gl}(3)$ \cite{Rowe1999,deGuise2020}. We can also combine these diagonal operators into traceless combinations which span the Cartan subalgebra of $\mathfrak{su}(3)$
\begin{align}
    \hat h_1&=\hat S_{11}-\hat S_{22}=2\hat S_z^{\{12\}},\\
    \hat h_2&=\hat S_{22}-\hat S_{33}=2\hat S_z^{\{23\}}.
\end{align}
Using these Cartan operators, we can label different SU(3) irreps by the eigenvalues $\lambda$ and $\mu$ of $\hat h_1$ and $\hat h_2$ for the highest weight state (h.w.s.), the state which is annihilated by all three raising operators. The highest weight state is guaranteed to be nondegenerate in these eigenvalues, and therefore the labels $(\lambda,\mu)$ uniquely specify the irreps of SU(3). These are known as \textit{Dynkin indices} \cite{Georgi1999}, and several examples of the weight spaces for particular Dynkin indices can be found in Fig. \ref{fig:su3_irreps}. The occupation numbers relate to these irrep labels as \cite{Martins2019,deGuise2020}
\begin{equation}
    \nu_1+\nu_2+\nu_3=\lambda+2\mu.\label{eq:occupation_sum}
\end{equation}
Equation (\ref{eq:occupation_sum}) tells us that the occupation numbers are not independent, which leads to multiplicities in states when labeled by the occupation numbers alone. To uniquely identify states in SU(3), we need to include one more label to lift this degeneracy. A common choice is to use the SU(2)$_{\{23\}}$ irrep label $I$. This is essentially the ``total angular momentum'' quantum number of a particular SU(2)$_{\{23\}}$ irrep within $(\lambda,\mu)$. Using the Dykin indices, occupation numbers, and $I$, we can write a state in an SU(3) irrep \cite{Rowe1999} as 
\begin{equation}
    \ket{(\lambda,\mu);(\nu_1,\nu_2,\nu_3)I}\equiv\ket{(\lambda,\mu);\boldsymbol{\nu}I},
\end{equation}
where $\boldsymbol{\nu}=(\nu_1,\nu_2,\nu_3)$. In cases where the SU(3) irrep is implied/defined we can drop the $(\lambda,\mu)$ label. Further, a primary focus of this work will be on $(\lambda,0)$ irreps, such as the $(9,0)$ irrep in Fig. \ref{fig:su3_irreps}, which has no multiplicities in the eigenvalues of $\hat h_1$ and $\hat h_2$. In this case we can uniquely label all states by the weight $(\nu_1-\nu_2,\nu_2-\nu_3)$, and $I$ becomes redundant.

With representations of SU(3) established, we can describe more complex objects like rotation operators, SU(3) Wigner-$D$ matrices, and SU(3) spherical tensor operators. In SU(2), arbitrary operators can be constructed by employing Euler angles
\begin{equation}
    \hat R(\alpha,\beta,\gamma)=e^{-i\alpha \hat J_z}e^{-i\beta \hat J_y}e^{-i\gamma \hat J_z},
\end{equation}
which allows one to describe an arbitrary element of the SU(2) Lie group using $(\alpha,\beta,\gamma)$. In quantum mechanics, global phases can be ignored, allowing one to instead consider the coset space SU(2)/U(1), where a U(1) symmetry has been ``modded out'' (the global phase). In such cases, one can consider the application of $\hat R(\alpha,\beta,0)$ on a fixed weight state of SU(2) (usually chosen to be the highest weight state $\ket{J,J}$) and thus describe all points in the SU(2)/U(1) manifold using $(\alpha,\beta)\equiv(\phi,\theta)$, the azimuthal and polar angles on the Bloch sphere.

In SU(3) one can make use of the three SU(2) subgroups to once again employ Euler angles to describe points in the manifold. A general SU(3) rotation operator can be written as \cite{Rowe1999,Klimov_SUn}
\begin{align}
    \nonumber\widetilde{\mathbf{R}}&(\alpha_1,\beta_1,\alpha_2,\beta_2,\alpha_3,\beta_3,\gamma_1,\gamma_2)\\
    \nonumber=&\hat R_{\{23\}}(\alpha_1,\beta_1,-\alpha_1) \hat R_{\{12\}}(\alpha_2,\beta_2,-\alpha_2)\\
    &\times\hat R_{\{23\}}(\alpha_3,\beta_3,-\alpha_3) e^{-i\gamma_1\hat h_1}e^{-i\gamma_2\hat h_2}\label{eq:full_rotation_operator}
\end{align}
with eight real parameters. The operators $\hat R_{\{ij\}}$ are the SU(2) rotation operators in subgroup $\{ij\}$ written in their Euler angle decomposition, and the final two operators in Eq.~(\ref{eq:full_rotation_operator}) are rotations generated by the Cartan operators. 

Similar to the case of SU(2), points on the manifold can be defined by fewer coordinates by appropriate choice of fiducial state. In SU(2), the manifold is parameterized by three parameters, but by choosing a fiducial state which is an eigenstate of the Cartan operator we reduce the manifold to one which contains only a polar and azimuthal coordinate, the manifold of SU(2)/U(1). In SU(3), one can similarly choose a fiducial state which is a simultaneous eigenstate of both Cartan operators, leading to a manifold with only six parameters
\begin{align}
    \nonumber\mathbf{R'}&(\alpha_1,\beta_1,\alpha_2,\beta_2,\alpha_3,\beta_3)\\
    \nonumber&\hspace{1.5cm}=\hat R_{\{23\}}(\alpha_1,\beta_1,-\alpha_1) \hat R_{\{12\}}(\alpha_2,\beta_2,-\alpha_2)\\
    &\hspace{2cm}\times\hat R_{\{23\}}(\alpha_3,\beta_3,-\alpha_3),\label{eq:rotation_6_parameters}
\end{align}
defining points on the manifold of SU(3)/U(2) \cite{Martins2019}.

A final simplification can be made by assuming that the manifold corresponds to a representation of the form $(\lambda,0)$, for example, the $(9,0)$ irrep in Fig. \ref{fig:su3_irreps}. Choosing the fiducial state in this irrep to be the highest weight state $\ket{\text{h.w.s.}}$ (represented in the figure with a red circle) means that rotations in the $\{23\}$ subgroup only apply a phase to the fiducial state. This same approach was taken in \cite{Klimov2017}. Therefore, the $\hat R_{\{23\}}$ operator in Eq.~(\ref{eq:rotation_6_parameters}) can be neglected, leading to a manifold defined by only four coordinates, $\Omega=(\alpha_1,\beta_1,\alpha_2,\beta_2)$ and a fiducial state. In the language of \cite{Biff_and_Mann}, this fiducial state is our coherent state at the origin of our coordinate system, and all points on the manifold $\Omega$ can be defined as
\begin{align}
    \ket{\Omega}\equiv& \hat R_{23}(\alpha_1,\beta_1,-\alpha_1)\hat R_{12}(\alpha_2,\beta_2)\ket{\text{h.w.s}}\\
    \equiv&\mathbf{R}(\Omega)\ket{\text{h.w.s.}}\label{eq:point_on_manifold}
\end{align}
Using the rotation operator in Eq.~(\ref{eq:point_on_manifold}) we can define our SU(3)/U(2) Wigner-D matrix as
\begin{equation}
    \hat D^{(\lambda,\mu)}_{\boldsymbol{\nu}'I';\boldsymbol{\nu}I}(\Omega)=\mel{(\lambda,\mu)\boldsymbol{\nu}'I'}{\mathbf{R}(\Omega)}{(\lambda,\mu)\boldsymbol{\nu}I}
\end{equation}
which is given by the matrix elements of $\mathbf{R}(\Omega)$ for irrep $(\lambda,\mu)$ between weight states $(\boldsymbol{\nu}I)$ and $(\boldsymbol{\nu}'I')$. 

The Clebsch-Gordan (CG) coefficients of $\mathfrak{su}(3)$ are defined as \cite{Martins2019}
\begin{align}
    \nonumber&\mathbf C_{(\lambda_1,\mu_1)\boldsymbol{\nu}_1I_1\;,\;(\lambda_2,\mu_2)\boldsymbol{\nu}_2I_2}^{(\lambda,\mu)\boldsymbol{\nu}I}\\
    &\hspace{0.5cm}\equiv\Big[\bra{(\lambda_1,\mu_1)\boldsymbol{\nu}_1I_1}\otimes\bra{(\lambda_2,\mu_2)\boldsymbol{\nu}_2I_2}\Big]\ket{(\lambda,\mu)\boldsymbol{\nu}I}.\label{eq:su3_cg}
\end{align}
Using this, we can write the SU(3) spherical tensor operators in $(\lambda,0)$ as \cite{Klimov_SUn,deGuise2020,Klimov2008,Martins2019}
\begin{align}
    \nonumber \hat T^{\lambda}_{\sigma,\boldsymbol{\nu}I}=\sum_{\boldsymbol{\nu'},\boldsymbol{\nu''}}(-1)^{\lambda-\nu''_2}&\mathbf C^{(\sigma,\sigma)\boldsymbol{\nu}I}_{(\lambda,0)\nu'I'\;,\;(0,\lambda)\left[\nu''I''\right]^*}\\
    &\times\ketbra{(\lambda,0);\boldsymbol{\nu'}I'}{(\lambda,0);\boldsymbol{\nu''}I''}\label{eq:su3_tensor}
\end{align}
where the $\left[\;\cdot\;\right]^*$ indicates the conjugate weight in the conjugate irrep $(0,\lambda)$. For irreps of type $(\lambda,0)$ and $(0,\lambda)$, which have no degeneracy, conjugating the weight simply maps $\nu_k\to\lambda-\nu_k$. Further, the lack of degeneracy in these irreps means the label $I$ is redundant, and weights can be determined from $\boldsymbol{\nu}$ alone; hence, we write the sum in Eq.~(\ref{eq:su3_tensor}) as running over only the occupation numbers $\boldsymbol{\nu'}$ and $\boldsymbol{\nu}''$.

\subsection{Comparisons with SU(2)}
\label{sec:heuristics}

Before moving on, we would like to offer the reader a few parallels between SU(3) and SU(2). The purpose of this is to provide an aid for understanding the objects defined in previous sections for those who are more familiar with SU(2).

First, recall that SU(2) irreps $\mathcal H_J$ of size $J$ can be constructed from tensor products of the fundamental representation $\mathcal H_{1/2}$ of SU(2) with itself according to Eq.~(\ref{eq:irrep_decomp}). Similarly, SU(3) irreps of types $(\lambda,\mu)$ appear in direct sum decompositions of tensor products of the fundamental representation $\mathcal{D}^{(1,0)}$ of SU(3) [although the decomposition is not as simple as Eq.~(\ref{eq:irrep_decomp})]. In general, $(\lambda,\mu)$ may appear with some multiplicity in the direct sum decomposition of $\otimes_{n=1}^N\mathcal{D}^{(1,0)}$. The exception to this is this tensor product space's symmetric subspace, $(N,0)$, which occurs only once. This is analogous to how the symmetric subspace $\mathcal{H}_{N/2}$ of $N$ spin-1/2 particles occurs only once in the direct sum decomposition in Eq.~(\ref{eq:irrep_decomp}).

In Sec. \ref{sec:collective_states} we defined angular momentum eigenstates using their total angular momentum $J$ and projective angular momentum $M$. However, one could instead label SU(2) irrep states by their occupation numbers, similar to our construction of the SU(3) weight state labels in Sec. \ref{sec:su3}. In this case, the occupation numbers are the number of particles up $\nu_\uparrow$, and the number of particles down $\nu_\downarrow$. For an irrep of size $J$ we can label states uniquely as
\begin{equation}
    \ket{J,M}\to\ket{J;(\nu_\uparrow,\nu_\downarrow)}
\end{equation}
where $M=(\nu_\uparrow-\nu_\downarrow)/2$. However, it's important to realize a key difference between SU(2) and SU(3) irreps when labeling states using occupation numbers. In SU(3) the occupation numbers of a given weight state are determined by the irrep type $(\lambda,\mu)$, whereas in SU(2) occupation numbers are \textit{not} uniquely determined by the irrep type $J$. For example, an SU(2) irrep of size $J$ has a highest weight state of $M=J$, but there are infinitely many possibilities for the occupation numbers $\nu_\uparrow$ and $\nu_\downarrow$ which correspond to this state. In SU(3), occupation numbers uniquely determine the highest weight state of a given irrep, and thus are a natural choice for labeling irreps $(\lambda,\mu)$.

Finally, in SU(2), spherical harmonics $Y_{k,q}$ are a special case of Wigner-$D$ matrices, being the matrix elements which couple the zero weight state to an arbitrary weight state within an irrep \cite{Sakurai2011}
\begin{equation}
    \mel{J,M}{\hat R(\alpha,\beta,0)}{J,0}=\sqrt{\frac{4\pi}{2J+1}}Y_{J,M}^*(\beta,\alpha).
\end{equation}
These functions play an important role in the SU(2) Wigner function kernel. Similarly, as we will demonstrate in the upcoming sections, the SU(3) analog of the spherical harmonics is the rotation operator [Eq.~(\ref{eq:point_on_manifold})] matrix element coupling the zero weight state of an irrep $(\sigma,\sigma)$ to an arbitrary weight $\ket{(\sigma,\sigma);\boldsymbol{\nu}I}$ within that irrep, and thus this function will play a similar role to the spherical harmonics in the SU(3) kernel.

\section{Isomorphism Between the Collective State Basis and SU(3) Irreps}
\label{sec:isomorphism}

In this section we establish an isomorphism between the vectorized collective state space of $N$ spin-1/2 particles [Eq.~(\ref{eq:vectorized_collective_state_space})] and an SU(3) irrep of type $(N,0)$. Pictorially, the mapping between weight diagrams is made by simply overlapping $\mathcal V$ with particle number $N$ (Fig.~\ref{fig:coll_space}) with an SU(3) irrep of type $(N,0)$ (Fig.~\ref{fig:su3_irreps}). This leads to a mapping between state labels
\begin{subequations}
\begin{align}
    \lambda&=N,\\
    \nu_1&=J+M,\\
    \nu_2&=J-M,\\
    \nu_3&=N-2J,\\
    I&=(N-J-M)/2.
\end{align}
\end{subequations}
Notice how, as discussed in Sec.~\ref{sec:su3}, only three of these numbers are necessary to define a state in the SU(3) irrep $(\lambda,0)$. Equivalently, only three quantum numbers ($J$, $M$, and $N$) are required to label a state in the collective state space.

Next, we can map raising and lowering operators in SU(3) to sums across generalized spherical tensor operators [Eq.~(\ref{eq:sph_tensor})] acting on the collective state space, similar to \cite{Forbes2024,Zhang2018}. Explicitly, we find that
\begin{subequations}
    \begin{align}
        \hat S_{12}&=-\sum_J\alpha(J)\hat T_{1,1}^{J,J}\equiv \hat J_+\\
        \hat S_{21}&=\sum_J\alpha(J)\hat T_{1,-1}^{J,J}\equiv \hat J_-\\
        \hat S_z^{\{12\}}&=\frac{1}{\sqrt2}\sum_J\alpha(J)\hat T_{1,0}^{J,J}\equiv \hat J_z\\
        \hat S_{13}&=\sum_J\beta(J,N)\hat T_{1/2,1/2}^{J+1/2,J}\\
        \hat S_{31}&=\sum_J\beta(J,N)\hat T_{1/2,-1/2}^{J,J+1/2}\\
        \hat S_{23}&=\sum_J\beta(J,N)\hat T_{1/2,-1/2}^{J+1/2,J}\\
        \hat S_{32}&=-\sum_J\beta(J,N)\hat T_{1/2,1/2}^{J,J+1/2}\\
        \hat Y&=\sum_J\gamma(J,N)\hat T_{0,0}^{J,J},
    \end{align}\label{eq:all_cs}
\end{subequations}
where
\begin{subequations}
    \begin{align}
        \alpha(J)&=\sqrt{\frac{2}{3}}\sqrt{J(J+1)(2J+1)},\\
        \beta(J,N)&=\sqrt{(N-2J)(J+1)(2J+1)},\\
        \gamma(J,N)&=\sqrt{2J+1}\left(2J-\frac{2}{3}N\right).
    \end{align}
\end{subequations}
The operator $\hat Y$, usually referred to as the \textit{hypercharge}, is the representation of the Gell-Mann matrix \cite{GellMann1962}
\begin{equation}
    \hat\lambda_8=\frac{1}{\sqrt3}\begin{pmatrix}
    1&0&0\\
    0&1&0\\
    0&0&-2
    \end{pmatrix}.
\end{equation}

The Eqs.~(\ref{eq:all_cs}) are written in terms of spherical tensor operators, but it is also convenient to look at these operators expressed as a sum across elements of $\mathcal B(\mathcal{V})$
\begin{subequations}
    \begin{align}
        \nonumber\hat S_{12}&=\sum_{J,M}\sqrt{(J-M)(J+M+1)}\\
        &\hspace{1cm}\times\ketbra{J,M+1,N}{J,M,N}\\
        \nonumber\hat S_{21}&=\sum_J\sqrt{(J+M)(J-M+1)}\\
        &\hspace{1cm}\times\ketbra{J,M-1,N}{J,M,N}\\
        \nonumber \hat S_{13}&=\sum_{J,M}\sqrt{(N-2J)(J+M+1)}\\
        &\hspace{1cm}\times\ketbra{J+1/2,M+1/2,N}{J,M,N}\\
        \nonumber \hat S_{31}&=\sum_{J,M}\sqrt{(N-2J+1)(J+M)}\\
        &\hspace{1cm}\times\ketbra{J-1/2,M-1/2,N}{J,M,N}\\
        \nonumber \hat S_{23}&=\sum_{J,M}\sqrt{(N-2J)(J-M+1)}\\
        &\hspace{1cm}\times\ketbra{J+1/2,M-1/2,N}{J,M,N}\\
        \nonumber \hat S_{32}&=\sum_{J,M}\sqrt{(N-2J+1)(J-M)}\\
        &\hspace{1cm}\times\ketbra{J-1/2,M+1/2,N}{J,M,N}\\
        \hat S_z^{\{12\}}&=\sum_{J,M}M\ketbra{J,M}{J,M}\\
        \hat Y&=\sum_{J,M}\left(2J-\frac{2}{3}N\right)\ketbra{J,M,N}.
    \end{align}\label{eq:ops_in_jm}
\end{subequations}
One can find the representations of all Gell-Mann matrices by taking linear combinations [Eq.~(\ref{eq:plus_minus})] of these ladder operators in Eqs.~(\ref{eq:all_cs}). Thus, the collective state space, together with these operators, is isomorphic to the $(N,0)$ irrep of SU(3).

\section{The Solid Spin Wigner Function}
\label{sec:wigner_function}

In this section we will make use of the isomorphism demonstrated in Sec.~\ref{sec:isomorphism} to construct an SU(3) Wigner function \footnote{The manifold of the Wigner function is formally SU(3)/U(2), but we refer to it as the SU(3) Wigner function analogous to how the SU(2)/U(1) Wigner function is referred to as the SU(2) Wigner function.} to represent operators on collective state space of size $N$. We will make use of some unique properties of collective state space to simplify the Wigner function to take only three real parameters as inputs. These parameters are interpreted as a polar, azimuthal, and radial component, thus allowing one to visualize the Wigner function on a Bloch ``ball'' rather than a Bloch sphere. We therefore refer to it as the \emph{solid} spin Wigner function. We will further show that the solid spin Wigner function can be reduced to an expression containing SU(2) spherical tensors and spherical harmonics. We will also drop the label $N$ from states in $\mathcal V$ from here on, as it is assumed we are working with $N$ spin-1/2 particles (i.e. $\ket{J,M,N}\to\ket{J,M}$).

\subsection{Defining the kernel}
In the following we define an SU(3) Wigner function using the construction in ref. \cite{Klimov_SUn}. We will construct the Wigner functions for operators on $(\lambda,0)$, where $\lambda=N$ following the isomorphism from the vectorized collective state space $\mathcal V$. The general form of the kernel $\hat\omega(\Omega)$ for an SU(3) Wigner function on representation $(\lambda,0)$ is
\begin{equation}
    \hat\omega(\Omega)=\sum_\sigma F_\sigma^\lambda \sum_{\boldsymbol{\nu},I} D_{\boldsymbol{\nu}I;\boldsymbol{0}0}^{\sigma}(\Omega)\hat T^{\lambda}_{\sigma;\boldsymbol{\nu}I}\label{eq:su3_kernel_general}
\end{equation}
where $F_\sigma^\lambda$ is a normalization factor, $\hat T^{\lambda}_{\sigma;\boldsymbol{\nu}I}$ is an SU(3) irreducible tensor operator, and $D_{\boldsymbol{\nu}I;\boldsymbol{0}0}^{\sigma}(\Omega)$ is a ``Wigner-$D$'' matrix element between the zero weight state in an irrep $(\sigma,\sigma)$ and $\boldsymbol{\nu}I$ (defined in Sec. \ref{sec:su3}) \cite{Martins2019,Klimov_SUn}. The factor $F_\sigma^\lambda$ can be found by enforcing normalization of $\hat\omega(\Omega)$ which implies that
\begin{align}
    F_\sigma^\lambda=\sqrt{\frac{2(\sigma+1)^3}{(\lambda+1)(\lambda+2)}}.
\end{align}

\begin{figure}
    \centering
    \includegraphics[width=0.94\linewidth]{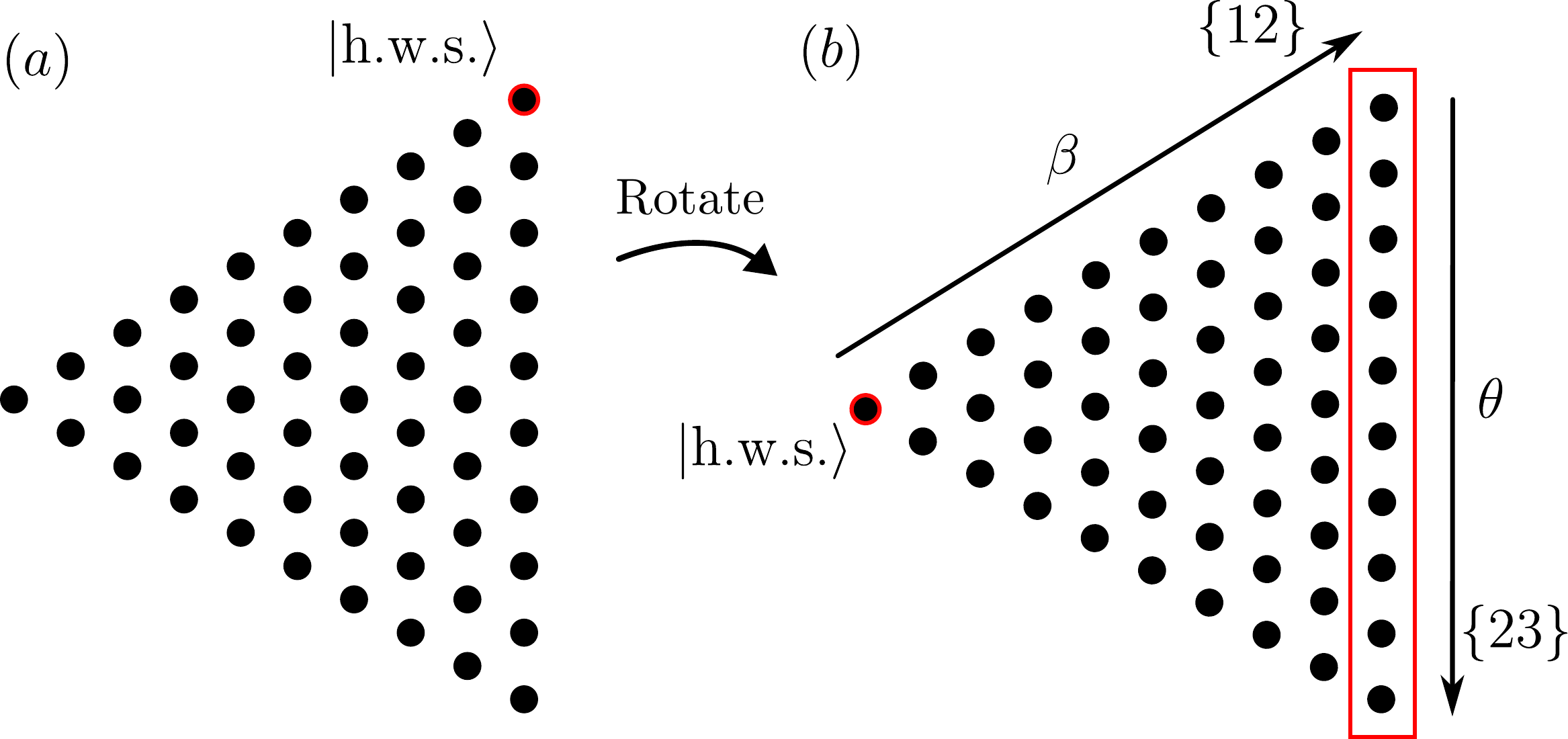}
    \caption{Diagram representing how the $(9,0)$ irrep (a) is used to encode the vectorized collective state space $\mathcal V$ (b). We highlight the highest weight state $\ket{\text{h.w.s.}}$ using a red circle, and note that in $\mathcal V$ it gets mapped to the $\ket{0,0}$ state. The symmetric subspace of $\mathcal V$, where $J=N/2$, is highlighted in red. The arrow corresponding to $\beta$ shows the direction that rotations by $\beta$ move the h.w.s. (in the $\{12\}$ subgroup), and similarly the arrow corresponding to $\theta$ shows how rotations by $\theta$ move the state (in the $\{23\}$ subgroup).}
    \label{fig:rotation_diagram}
\end{figure}

Using the isomorphism in Sec. \ref{sec:isomorphism}, we can embed collective states in an SU(3) irrep. We have a choice of which $\mathfrak{su}(2)_{\{ij\}}$ subalgebra within $(\lambda,0)$ to use to represent the $\mathfrak{su}(2)$ algebra of $\mathcal C$, and we will choose the $\{23\}$ subalgebra. Figure \ref{fig:rotation_diagram} shows an example of how the $N=9$ collective state space is represented using the $(9,0)$ SU(3) irrep. The choice of the $\{23\}$ subgroup to represent the SU(2) group of $\mathcal C$ can be thought of as rotating the SU(3) irrep so that the highest weight state is mapped to the $\ket{J=0,M=0,N}$ state in $\mathcal V$. The symmetric subspace of $\mathcal V$ is highlighted in red, and is embedded in the $\{23\}$ subalgebra. We choose this convention for several reasons, one being mathematical convenience due to the fact that the degeneracy label in weight states of SU(3) is usually chosen to be the SU(2)$_{\{23\}}$ total angular momentum $I$, as discussed in Sec. \ref{sec:su3}. Another reason why this choice of convention is useful here is that it leads to an easier interpretation of the topology of the manifold as a solid ball, as we will come to see. Note that this choice of convention is a rotated version of the convention used to describe the operators in Eq.~(\ref{eq:ops_in_jm}), where the $\{12\}$ subgroup was chosen to represent the $\mathfrak{su}(2)$ algebra in $\mathcal C$.

Using this convention we can write, by inspection of the diagram in Fig. \ref{fig:rotation_diagram}, that for $(\lambda,0)$ irreps $2M=\nu_2-\nu_3$, $J=I$, and $N=\lambda$. We will write the SU(3) Wigner function using these parameters. We first note that $\mathfrak{su}(3)$ CG coefficients can be written as an $\mathfrak{su}(2)$ CG coefficient, multiplied by a reduced matrix element \cite{deGuise2020}. By considering the form of the $\mathfrak{su}(3)$ CG coefficient in Eq.~(\ref{eq:su3_tensor}), and enforcing that $I'=I''=J$, one can write
\begin{equation}
   \mathbf C_{(N,0)\boldsymbol{\nu}'J\;,\;(0,N)\boldsymbol{\nu}''J}^{(\sigma,\sigma)\boldsymbol{\nu}k}=A_{J,N}^{\sigma,k}C_{J,M;J,-(M+q)}^{k,-q}\label{eq:su3_to_su2_cg}
\end{equation}
where $C_{J,M;J,-(q+M)}^{k,-q}$ is a regular $\mathfrak{su}(2)$ CG coefficient and $A_{J,N}^{\sigma,k}$ is a reduced coefficient which can be found explicitly in Eq.~(\ref{eq:A_reduced}) in App. \ref{sec:reducing_A}. In Eq.~(\ref{eq:su3_to_su2_cg}) we have defined $q=(\nu_3-\nu_2)/2$, $M=(\nu'_2-\nu'_3)/2$, and $-(q+M)=(\nu''_2-\nu''_3)/2$ (required by triangularity of CG coefficients). Substituting Eq.~(\ref{eq:su3_to_su2_cg}) into Eq.~(\ref{eq:su3_tensor}) we obtain a simplified expression for the SU(3) irreducible tensor operator
\begin{align}
    \nonumber \hat T^N_{\sigma,\boldsymbol{\nu}J}&=\sum_{J,M,q}(-1)^{J+M+q}A_{N,J}^{\sigma,k}C^{k,-q}_{J,M;J,-(M+q)}\\
    &\hspace{3cm}\times\ketbra{J,M}{J,M+q}\\
    &=\sum_{J,q}(-1)^{2J+q}A_{N,J}^{\sigma,k}\left(\hat T_{k,q}^{(J)}\right)^\dagger
\end{align}
where $\hat T_{k,q}^{(J)}$ is an SU(2) spherical tensor operator acting on an irrep of size $J$. For details on this calculation see App.~\ref{sec:reducing_A}.

In Eq.~(\ref{eq:su3_to_su2_cg}) we enforced that the $\mathfrak{su}(2)$ CG coefficient only couples irreps of equal total angular momentum. This enforces that $\nu_1$, the first occupation number in $(\sigma,\sigma)$, be $\nu_1=\sigma$ \cite{Martins2019,deGuise2020}. This results in a very important simplification to the SU(3) Wigner-$D$ matrix in Eq.~(\ref{eq:su3_kernel_general}). From \cite{Klimov2017} we can write
\begin{align}
    \nonumber D^\sigma_{\boldsymbol{\nu}k;\boldsymbol{0}0}=&(-1)^\sigma\sqrt{2k+1}D^{k}_{\frac{1}{2}(\nu_2-\nu_3),\frac{1}{2}(\sigma-\nu_1)}(\alpha_1,\beta_1,-\alpha_1)\\
    \nonumber &\times\sum_{J'}(-1)^{J'}\frac{2J'+1}{\sigma+1}\left\{\begin{array}{ccc}
        \sigma/2 & (2\sigma-\nu_1)/2 & k  \\
         \nu_1/2 & \sigma/2 & J' 
    \end{array}\right\}\\
    &\times D^{J'}_{\nu_1-\sigma,0}(\alpha_2,\beta_2,-\alpha_2)\label{eq:D_unsimplified}
\end{align}
where $(\alpha_1,\beta_1,\alpha_2,\beta_2)$ are the group parameters from Eq.~(\ref{eq:point_on_manifold}) and $D^k_{q',q}$ are elements of the SU(2) Wigner-$D$ matrix. Substituting $\nu_1=\sigma$ and $q=(\nu_3-\nu_2)/2$ this becomes
\begin{align}
    \nonumber D^\sigma_{qk;\boldsymbol{0}0}=&(-1)^\sigma\sqrt{2k+1}D^{k}_{-q,0}(\alpha_1,\beta_1,-\alpha_1)\\
    \nonumber &\times\sum_{J'}(-1)^{J'}\frac{2J'+1}{\sigma+1}\left\{\begin{array}{ccc}
        \sigma/2 & \sigma/2 & k  \\
         \sigma/2 & \sigma/2 & J' 
    \end{array}\right\}\\
    &\times D^{J'}_{0,0}(\alpha_2,\beta_2,-\alpha_2)\\
    \nonumber=&\frac{4\pi(-1)^\sigma}{\sigma+1} (-1)^qY^k_q(\theta,\phi)\sum_{J'}(-1)^{J'}\sqrt{2J'+1}\\
    &\hspace{1.7cm}\times\left\{\begin{array}{ccc}
        \sigma/2 & \sigma/2 & k  \\
         \sigma/2 & \sigma/2 & J' 
    \end{array}\right\}Y^{J'}_0(\beta,0),\label{eq:D_simplified}
\end{align}
where in the last line we have replaced the Wigner-$D$ matrix elements with their equivalent expressions in terms of spherical harmonics $Y^k_q(\theta,\phi)$, and have introduced a new notation $\theta\equiv\beta_1$, $\phi\equiv\alpha_1$, and $\beta\equiv\beta_2$. Notice that in the last line Eq.~(\ref{eq:D_simplified}) the azimuthal angle $\alpha_2$ in the second spherical harmonic has been removed since the spherical harmonic has 0 projection, and thus is only a function of its polar angle which we define as $\beta$.

Combining our simplified expressions for the SU(3) irreducible tensors and Wigner-$D$ matrix, we can write the kernel from Eq.~(\ref{eq:su3_kernel_general}) explicitly as
\begin{align}
    \nonumber\hat\omega(\theta,\phi,\beta)=&4\pi\sum_\sigma F_{\sigma}^N\sum_{J,J',k,q}\frac{(-1)^{\sigma+2J+J'}}{\sigma+1}A_{N,J}^{\sigma,k}\\
    \nonumber&\times\sqrt{2J'+1}\left\{\begin{array}{ccc}
        \sigma/2 & \sigma/2 & k  \\
         \sigma/2 & \sigma/2 & J' 
    \end{array}\right\}Y^{J'}_0(\beta)\\
    &\times Y^k_q(\theta,\phi)\left(\hat T_{k,q}^{(J)}\right)^\dagger.\label{eq:su3_kernel_simplified}
\end{align}

\subsection{Interpreting the kernel}

The final expression we have derived for the kernel, Eq.~(\ref{eq:su3_kernel_simplified}), contains only three angles, a significant reduction from the eight real parameters needed to define the full SU(3) manifold. Since we chose to represent the $\mathfrak{su}(2)$ algebra of $\mathcal C$ using the $\{23\}$ subalgebra of $(\lambda,0)$, the angles $(\theta,\phi)$ can be interpreted as polar and azimuthal angles defining a point on a sphere for some fixed value of $\beta$. Since $(\theta,\phi)$ define rotations in the $\{23\}$ subgroup, that leaves $\beta$ to describe how the population of the state is distributed across the different SU(2) irreps in $\mathcal C$. The role of these parameters is demonstrated in Fig. \ref{fig:rotation_diagram}, where $\beta$ controls population across SU(2)$_{\{23\}}$ irreps.

\begin{figure}
    \centering
    \includegraphics[width=0.85\linewidth]{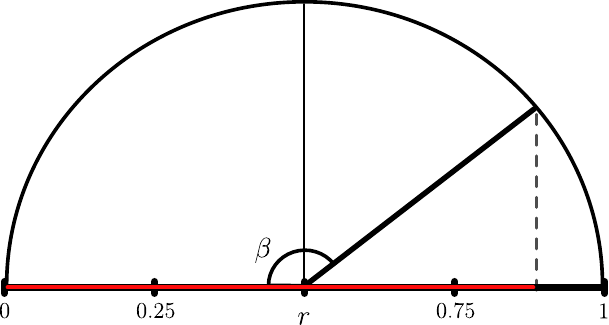}
    \caption{Diagram representing the change of variables in Eq.~(\ref{eq:beta_to_r}) from polar angle $\beta$ to radial component $r$.}
    \label{fig:radial_diagram}
\end{figure}

If we had not enforced that different SU(2) irreps are uncoupled, the kernel would retain an azimuthal angle in addition to the polar angle $\beta$ describing rotations within the SU(2)$_{\{12\}}$ subgroup. The fact that this angle becomes unnecessary intuitively comes from the fact that no phase information exists between SU(2) irreps in collective state space $\mathcal C$.

Another important note is that our choice of representing the $\mathfrak{su}(2)$ algebra of $\mathcal C$ within the $\mathfrak{su}(2)_{\{23\}}$ subalgebra of $(\lambda,0)$ means that the spin coherent state $\ket{N/2,N/2}\in\mathcal C$ is mapped to the $\ket{(0,N,0)N/2}$ state in $(N,0)$, which is \emph{not} the highest weight state $\ket{(N,0,0)0}$ as can be seen in Fig. \ref{fig:rotation_diagram}. This results in the coherent state at the point $(\theta,\phi,\beta)=(0,0,\pi)$ representing $\ket{N/2,N/2}$ as opposed to $(0,0,0)$ as one might have expected. This is simply a choice of convention, but lends to the interpretation of $\beta$ representing a ``radial'' component of the Wigner function.

\begin{figure*}
    \centering
    \includegraphics[width=1\linewidth]{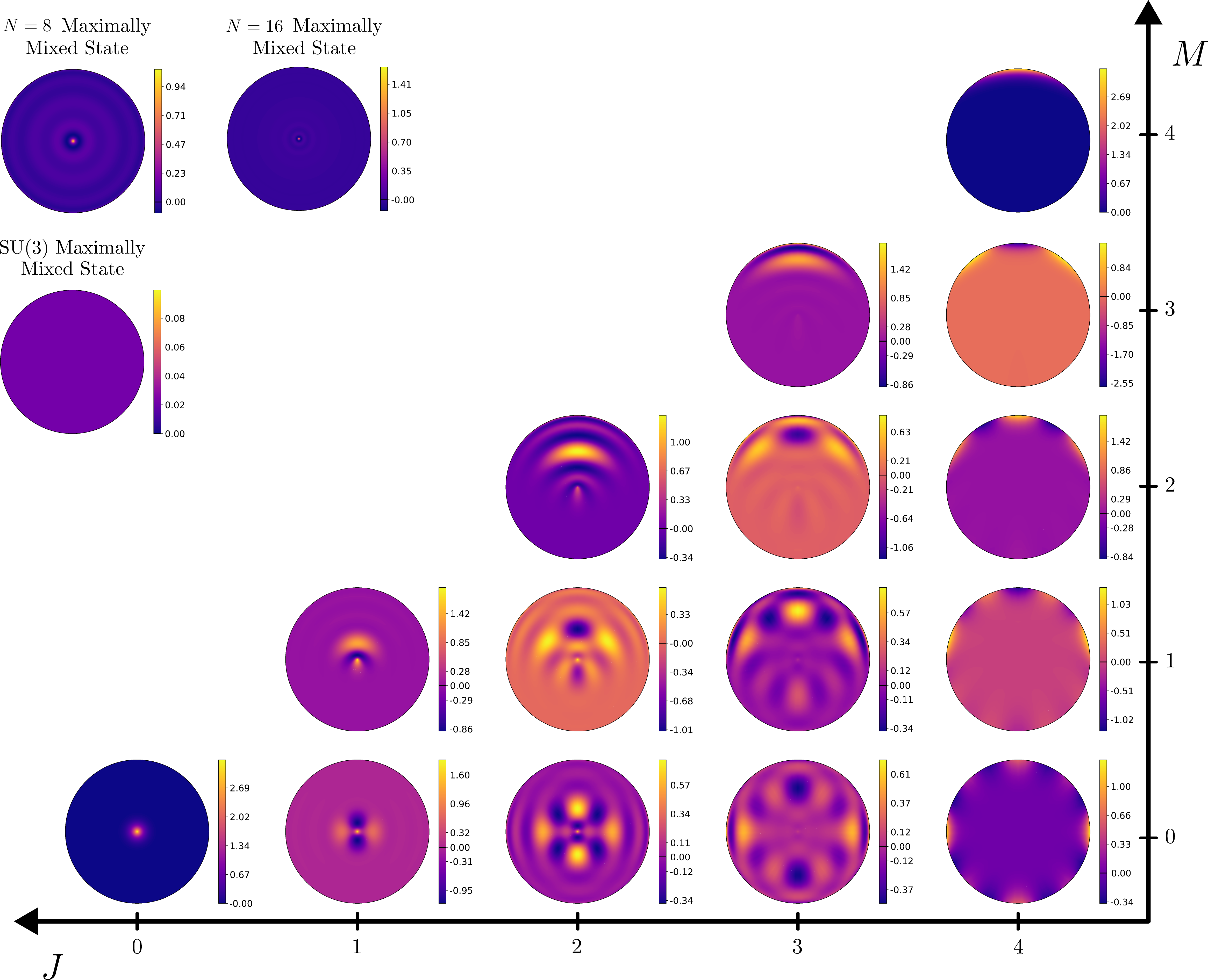}
    \caption{Slices of solid spin Wigner functions along $\phi=0,\pi$ for angular momentum eigenstates $\ketbra{J,M}$ for $N=8$. Each plot is normalized to its own minimum and maximum value, to ensure that the features of each plot are clearly visible. Also plotted are three examples of maximally mixed states. The $N=8$ and $N=16$ maximally mixed states are those defined by Eq.~(\ref{eq:maximally_mixed}), and are the maximally mixed states of $N$ spin-1/2 particles. Finally, the SU(3) maximally mixed state, Eq.~(\ref{eq:su3_maximally_mixed}), is plotted. This is the state corresponding to equal weight on all states in the $(N=8,0)$ irrep of SU(3), and its Wigner function is thus constant in its parameters.}
    \label{fig:dicke_states}
\end{figure*}

\subsection{Radial Form of the Kernel}
A convenient simplification to the measure of the manifold occurs when we make a change of variables stemming from the interpretation of $\beta$ as encoding information about a radial component of the Wigner function. Since $\beta$ is the ``polar'' angle in the SU(2)$_{\{12\}}$ subgroup, we define the radial component $r$ as the scaled projection of $\beta$ onto its corresponding $z$ axis (see Fig.~\ref{fig:radial_diagram})
\begin{equation}
    r=\frac{1-\cos(\beta)}{2}.\label{eq:beta_to_r}
\end{equation}
The measure $\mathrm d\Omega$ on the SU(3)/U(2) manifold, integrating over $\alpha_2$, is \cite{Martins2019,Byrd1997}
\begin{equation}
    \mathrm d\Omega=2\pi\sin(\theta)\cos(\beta/2)\sin(\beta/2)^3\mathrm d\theta\mathrm d\phi\mathrm d\beta,
\end{equation}
which, under the change of variables, becomes
\begin{equation}
    \mathrm d\Omega=2\pi\sin(\theta)r\mathrm dr\mathrm d\theta\mathrm d\phi.\label{eq:r_int_element}
\end{equation}

The above measure bears resemblance to that of a true solid ball, with the exception being a missing factor of $r$. Thus, the interpretation of the solid spin Wigner function as a ball is only a nice tool for visualization and simplification, but the manifold itself does not naturally share its geometry with that of a solid ball.

Transforming from $\beta$ to $r$ also allows us to rewrite the kernel as
\begin{align}
    \nonumber\hat\omega(\theta,\phi,r)=&\sqrt{4\pi}\sum_{\sigma,J,J',k,q}F_\sigma^N\frac{(-1)^{\sigma+2J+J'}}{\sigma+1}A^{\sigma,k}_{N,J}\\
    \times(2J'+1)&\begin{Bmatrix}
        \sigma/2 & \sigma/2 & k\\
        \sigma/2 & \sigma/2 & J'
    \end{Bmatrix}P_{J'}(1-2r)Y^k_q(\theta,\phi)^*\hat T_{k,q}^{(J)}\\
    =&\sqrt{4\pi}\sum_{J,k,q}R_{J,k}(r)Y^k_q(\theta,\phi)\left(\hat T_{k,q}^{(J)}\right)^\dagger,\label{eq:kernel_using_R}
\end{align}
where in the final line we have identified a function $R_{J,k}(r)$ called the \emph{radial function}, defined as
\begin{align}
    \nonumber R_{J,k}(r)=&\sum_{\sigma,J'}F_\sigma^NA^{\sigma,k}_{N,J}\frac{(-1)^{\sigma+J'+2J}}{\sigma+1}\\
    &\times(2J'+1)\begin{Bmatrix}
        \sigma/2 & \sigma/2 & k\\
        \sigma/2 & \sigma/2 & J'
    \end{Bmatrix}P_{J'}(1-2r).\label{eq:radial_component}
\end{align}
Equation~(\ref{eq:kernel_using_R}) has a similar structure to the standard SU(2) Wigner function, but with a factor of $R_{J,k}(r)$. This radial function depends on the Legendre polynomials, which we have substituted in place of the $\beta$-dependent spherical harmonic. The radial function for fixed $k$ must form an orthogonal family of functions indexed by $J'$ in order for traciality to be satisfied. From normalization of the Wigner function (see App.~\ref{app:normalization}) we deduce that
\begin{equation}
    \int_0^1R_{J,k}R_{J',k}r\mathrm dr=\frac{\delta_{J,J'}}{2d},
\end{equation}
where $d=2((N+1)(N+2))^{-1}$ is the dimension of $(N,0)$ \cite{Martins2019}. Further, unit trace of the kernel requires that
\begin{equation}
    \sum_J\sqrt{2J+1}R_{J,0}(r)=1.
\end{equation}

\section{Visualizations of the Solid Spin Wigner Wigner}
\label{sec:visualizations}

\begin{figure*}
    \centering
    \includegraphics[width=0.9\linewidth]{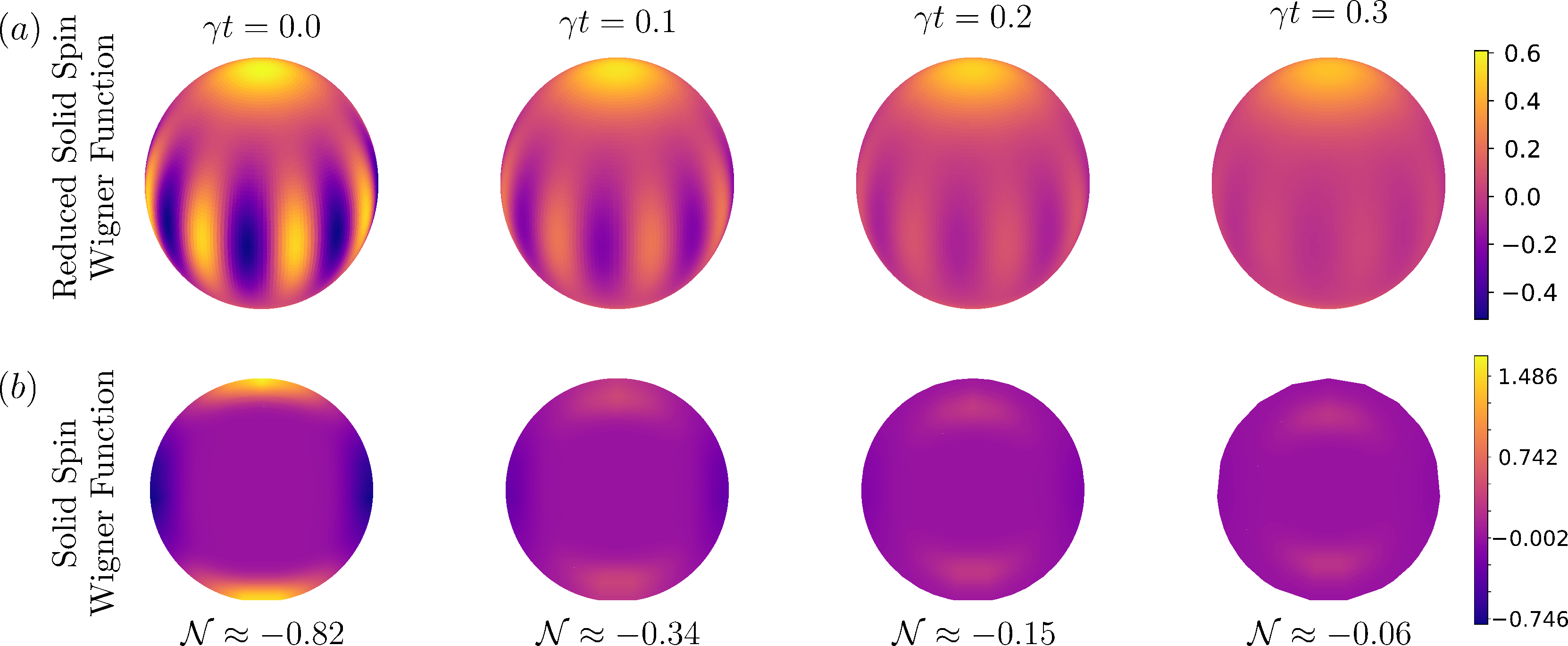}
    \caption{Snapshots of the reduced solid spin Wigner function (top) and a slice of the solid spin Wigner function (bottom) for an initial GHZ state $\ket\psi=(\ket{J,J}+\ket{J,-J})/\sqrt2$ undergoing atom loss at a rate of $\gamma$ per atom. The slice of the Wigner function is taken at $\phi=0,\pi$. The reduced Wigner function begins with well-defined features, which slowly get washed out as a function of the noise. Similarly, the slice of the Wigner function initially has negative regions on the sides where $\theta=\pi/2$, but as a function of time these negative regions disappear. Simultaneously, the positive lobes at the north and south poles drift inward towards the origin. The negativity $\mathcal N$ of each state is shown below its corresponding plot, with it decreasing as a function of the noise.}
    \label{fig:ghz_states}
\end{figure*}

We present here two methods of visualizing the solid spin Wigner function, each with particular use cases. The first involves taking a slice along a particular azimuthal angle $\phi$, and plotting the resulting function of $r$ and $\theta$. This can be thought of as slicing the solid ball in half and looking at the face which has been cut. The second method we will use is integration over the radial component, which leaves the resulting Wigner function dependent on only $\theta$ and $\phi$. We refer to this as the reduced solid spin Wigner function, and plot it on the surface of a sphere.

\subsection{Azimuthal Slice of Wigner Function}
\label{sec:slice}

For functions which are azimuthally symmetric, or for which the radial and polar components are of most interest, taking a slice of the Wigner function at a value of $\phi$ is most informative. For example, in Fig. \ref{fig:dicke_states}, angular momentum eigenstates for an $N=8$ system are shown, with the slice taken at $\phi=0$ (on the right half, and $\phi=\pi$ on the left half). In the figure, the radial distance from the center is $r$ and the polar angle $\theta$ runs from 0 at the top of the plot, to $\theta=\pi$ at the bottom.

This visualization highlights the radial distribution of the Wigner function. Interestingly, the Wigner functions of many states look similar to what one might have desired intuitively. For example, in Fig.~\ref{fig:dicke_states} all states with maximum $J=N/2$ have more weight towards the surface of the ball where $r=1$. Conversely, the lower $J$ states have more weight towards the center of the ball, with the $\ketbra{0,0}$ state being represented by a Wigner function with weight only near the center. The stretched states $\ket{J,J}$ of each SU(2) irrep have a distinct wide yellow region, which is at the top of the ball for the largest irrep, but moves down for each lower irrep. Finally, all states in the largest irrep have distributions around the surface of the ball which look similar to that of the SU(2) Wigner function. For example, the SU(2) Wigner function of a first-exited Dicke state $\ket{N/2,N/2-1}$ is a positive ring around the north pole, with a negative region in the center. This is exactly what is seen in the plot of $\ket{4,3}$ in Fig.~\ref{fig:dicke_states} , where the slice shows a cross section of this ring and the negative region.

\subsection{The Reduced Solid Spin Wigner Function}
\label{sec:reduced_wigner_function}

Another useful tool for visualizing the solid spin Wigner function is to integrate over the radial component and plot this ``marginalized'' Wigner function on a sphere. This function bears resemblance to the SU(2) Wigner function for states in the symmetric subspace, but becomes ``washed out'' under noise.

Integrating over $r$ in Eq.~(\ref{eq:radial_component}) can be computed explicitly thanks to properties of Legendre polynomials. We note that
\begin{equation}
    \int_{0}^1P_{J'}(1-2r)r\mathrm dr=\begin{cases}
        1/2 & J'=0\\
        -1/6 & J'=1\\
        0 & J'>1
    \end{cases},
\end{equation}
and thus the sum over $J'$ in Eq.~(\ref{eq:radial_component}) reduces to only two terms when integrating over $r$. Explicitly, one is left with
\begin{align}
    \nonumber \int_0^1 &R_{J,k}(r)r\mathrm dr=\frac{\sqrt{2J+1}}{2d}\delta_{k,0}\\
    &+\sum_{\sigma=\text{max}(k,1)}^N(-1)^{2J+k}\frac{k(k+1)}{\sigma(\sigma+1)^2(\sigma+2)}F_\sigma^N A_{N,J}^{\sigma,k}
\end{align}
where the first term is the case of $\sigma=k=0$, and $d$ is the dimension of the irrep $(N,0)$. Defining the above integral as $\gamma_{J,k}$ we write that the reduced Wigner function kernel $\hat\omega_{\text{red}}(\theta,\phi,r)$ is
\begin{equation}
    \hat\omega_{\text{red}}(\theta,\phi,r)=\sum_{J,k,q}\gamma_{J,k}Y_q^k(\theta,\phi) \left(\hat T_{k,q}^{(J)}\right)^\dagger,
\end{equation}
which bears similarity to the SU(2) Wigner function \cite{Agarwal1981}, albeit with a weighing function $\gamma_{J,k}$. This kernel is not a true Wigner function kernel, as it does not define a bijective Weyl map. However, it is convenient for visualizations of systems undergoing noise, as it can be represented on a sphere as seen in Fig. \ref{fig:ghz_states}. In the figure, atom loss noise is modeled similar to the jump formalism from \cite{Zhang2018} (but using direct master equation simulation by defining jump operators \cite{Forbes2024}) mapped to the $(N,0)$ irrep. In Fig. \ref{fig:ghz_states}(a) the reduced Wigner function is shown for different amounts of atom loss occurring at a rate $\gamma$ per atom. In Fig. \ref{fig:ghz_states}(b) the corresponding Wigner function slice is shown, with $\phi=0,\pi$ chosen as the slice. In both (a) and (b) the noise smooths out features of the state as a function of time. Further, in (b) the drift of the top and bottom ``lobes'' from the poles towards the center of the sphere is visible. The corresponding Wigner negativity $\mathcal N$ of each state is shown below the corresponding plot, and it decreases as a function of the noise.

\section{Discussion}
\label{sec:discussion}
While the kernel in Eq.~(\ref{eq:kernel_using_R}) satisfies the SW conditions, the mapping from $\mathcal C$ to $\mathcal B(\mathcal V)$, as well as the embedding of the state within an SU(3) irrep, leads to some unexpected behaviors of the solid spin Wigner function which should be discussed.

\subsection{Negativity}
\label{sec:negativity}
Unlike the Wigner negativity of bosonic modes, which has well-established operational meaning \cite{Albarelli2018,Mari2012,Veitch2012,Delfosse2015}, the Wigner negativity of SU(2) systems has a less clear interpretation (for continuous Wigner functions). For example, the spin coherent state $\ket{\uparrow}^{\otimes N}$, which is in a sense the ``most classical'' state of an ensembles of spin-1/2 particles, contains negativity. This negativity decreases as the system size grows, but its existence raises questions about the operational meaning of continuous-variable Wigner negativity in finite dimensional Hilbert spaces. Similarly, the solid spin Wigner function exhibits negativity in states that one might interpret as ``classical''. One reason for this negativity in the solid spin Wigner function is that the Wigner function assumes SU(3) group actions which do not natively exist in the underlying system.

The only physically relevant group actions on $\mathcal C$ are the collective SU(2) rotations, the same as for the usual SU(2) Wigner function. However, the space of $\mathcal C$ is very different from the single SU(2) irrep for which the SU(2) Wigner function is defined. In this work, we managed to define a Wigner function for $\mathcal C$ by assuming additional group actions associated with the SU(3) group, which leads to a larger representation space $(\lambda,0)$, isomorphic to $\mathcal V$. However, these additional group actions, associated with SU(2) rotations in $\{12\}$ and $\{13\}$ subgroups, do not naturally occur for states represented in $\mathcal C$ \footnote{If one were to define such group actions on $\mathcal C$, they would necessarily be local operations, not collective.}. As a result, the covariance property of the SW conditions is satisfied only for rotations within SU(2)$_{\{23\}}$, as no unitary exists in $\mathcal C$ which performs a rotation in the SU(2)$_{\{12\}}$ or SU(2)$_{\{13\}}$ subgroups. Despite this, the covariance property from the SW conditions is still satisfied by the kernel in Eq.~(\ref{eq:su3_kernel_simplified}), as the only group action we wish for covariance to apply to is the rotation in SU(2)$_{\{23\}}$. However, since the SU(3) Wigner function is constructed by assuming additional group actions than just those in SU(2)$_{\{23\}}$, the resulting Wigner function has negativity which may be questionable.

Through assuming no coupling between SU(2)$_{23}$ irreps of different $J$, we reduced these additional group actions to just a single one, the rotation by $\beta$ about the $\hat S_y^{\{12\}}$ axis,
\begin{equation}
    \hat U_\beta=e^{-i\beta\hat S_y^{\{12\}}}.
\end{equation}
Because this group action has no physically relevant implementation in $\mathcal C$, we observe negativity in the Wigner function across the radial component, which may not represent physically ``accessible'' negativity using collective operations. One may interpret this negativity across the radial component as arising from a necessity to satisfy the traciality condition. For this reason, the negativity of the solid spin Wigner function should not be taken at face value, and more research is needed to connect it to some notion ``quantumness'' as is often desired. Despite this, the spin coherent $\ket{N/2,N/2}$ has vanishing negativity as the system size increases, as expected. This can be seen by noting that the spin coherent state $\ket{N/2,N/2}$ is also a coherent state in the SU(3) manifold \cite{Biff_and_Mann}, as it is only an SU(2) rotation away from the highest weight state. Thus, since the highest weight state is expected to have vanishing negativity as $N\to\infty$, we should expect the same for $\ket{N/2,N/2}$.

\subsection{The Maximally Mixed State}
\label{sec:maximally_mixed_state}

The maximally mixed state is required to have a constant Wigner function by the unit trace condition of the kernel. However, the maximally mixed state of $N$ spin-1/2 particles, when mapped to the $(\lambda,0)$ irrep, is not the maximally mixed state of $(\lambda,0)$. Written in the collective state basis, the maximally mixed state $\hat\rho_{\text{M.M.}}$ of $N$ spin-1/2 particles is
\begin{align}
    \hat\rho_{\text{M.M}}&=\frac1{2^N}\sum_{J,M,\alpha}\ketbra{J,M,\alpha}\\
    &=\frac{1}{2^N}\sum_{J,M}d_J^N\col{J,M,N}{J,M,N},\label{eq:maximally_mixed}
\end{align}
where we see that the degeneracy weighs collective states. Further, the index $J$ in Eq.~(\ref{eq:maximally_mixed}) runs over integer values of $J$. In contrast, the maximally mixed state $\hat\rho_{\text{M.M.}}'$ in the SU(3) irrep $(N,0)$ is
\begin{equation}
    \hat\rho_{\text{M.M.}}'=\frac{1}{\frac{1}{2}(N+1)(N+2)}\sum_{J,M}\ketbra{J,M}\label{eq:su3_maximally_mixed}
\end{equation}
where $J$ runs over half integer steps. This difference means that the Wigner function for the maximally mixed state $\hat\rho_{\text{M.M.}}$ is not constant over the parameters $(\theta,\phi,r)$ as can be seen in Fig.~\ref{fig:dicke_states}. However, numerically it appears that the negativity of these maximally mixed states decreases as $N$ increases. By comparison, the SU(3) maximally mixed state from Eq.~(\ref{eq:su3_maximally_mixed}) is constant.

\subsection{Radial Coordinate Transformation}
\label{sec:coordinate_transformation}
Finally, we would like to point out that nontrivial identities for $R_{J,k}(r)$ and $A_{N,J}^{\sigma,k}$ can be derived by making use of the covariance property of the Wigner function. For example, one can write that for a group transformation $g_\pi$ represented by $\hat U_\pi^{\{12\}}$ which applies an SU(2) $\pi$-pulse to the SU(2)$_{\{12\}}$ subgroup, the Wigner function obeys
\begin{equation}
    \Tr(\hat U_\pi^{\{12\}}\hat\rho(\hat U_\pi^{\{12\}})^\dagger\hat\omega(\Omega))=\Tr(\hat\rho\hat\omega(g_\pi^{-1}\cdot\Omega)).
\end{equation}
The action of $g_\pi$ on $\Omega$ can be calculated by inspecting the result of applying $\hat U_\pi^{\{12\}}$ to the state $\ket\Omega$ which represents a coherent state at the point $\Omega$ on our manifold defined by Eq.~(\ref{eq:point_on_manifold}). We remind the reader that, after our simplification which assumed no coupling between SU(2)$_{\{23\}}$ irreps with differing $J$, we eliminated $\alpha_2$ from the definition of $\Omega$. Further, applying $\hat U_\pi^{\{12\}}$ to a state which contains coherence within a particular SU(2)$_{23}$ irrep will transform those coherences into coherences between irreps of differing $J$. Therefore, in order to represent $\hat U_\pi^{\{12\}}$ as a coordinate transformation, we must assume no coherence between weights within an SU(2)$_{23}$ irrep. This assumption implies that any states we wish to apply $\hat U_\pi^{\{12\}}$ to, and utilize the same 3-dimensional manifold $(\theta,\phi,r)$, must be $\phi$-independent, as $\phi$ encodes coherence between weights in the SU(2)$_{\{23\}}$ irreps. Thus, when considering the action of $g_\pi$ on $\Omega$ we are forced to assume the state is $\phi$-independent, leading to a manifold we need only define by $(\theta,r)$ or $(\theta,\beta)$.

Using this simplification, we write that
\begin{align}
    \hat U_\pi^{\{12\}}\ket{\theta,r}&=e^{-i\pi \hat S_y^{\{12\}}}e^{-i\theta \hat S_y^{\{23\}}}e^{-i\beta \hat S_y^{\{12\}}}\ket{0,0}\\
    &=\ket{\theta',r'},
\end{align}
where the goal is to solve for $\theta'$ and $r'$. This can be done simply by solving the equations in the fundamental representation, where the generators are Gell-Mann matrices \cite{GellMann1962}, and the highest weight state is $\ket{0,0}\equiv(1,0,0)^T$. Thus solving for $\theta'$ and $r'$ can be done by solving a system of three equations
\begin{equation}
    e^{-i\pi\hat\lambda_2/2}e^{-i\theta\hat\lambda_7/2}e^{-i\beta\hat\lambda_2/2}\begin{pmatrix}
        1\\
        0\\
        0
    \end{pmatrix}=e^{-i\theta'\hat\lambda_7/2}e^{-i\beta'\hat\lambda_2/2}\begin{pmatrix}
        1\\
        0\\
        0
    \end{pmatrix}
\end{equation}
where $\hat\lambda_2$ and $\hat\lambda_7$ are two of the Gell-Mann matrices given by
\begin{equation}
    \hat\lambda_2=\begin{pmatrix}
        0&-i&0\\
        i&0&0\\
        0&0&0
    \end{pmatrix}\;\;\;\;\;\text{and}\;\;\;\;\;\hat\lambda_7=\begin{pmatrix}
        0&0&0\\
        0&0&-i\\
        0&i&0
    \end{pmatrix}.
\end{equation}
The equation is rendered easier to solve by transforming $\theta$ to its linearized parameter $s$, identically to how $\beta$ was linearized to $r$, i.e. $s=(1-\cos(\theta))/2$. Solving these coupled equations, we find that
\begin{subequations}
\begin{align}
    r'&=1+r(s-1),\label{eq:coordinate_transformation1}\\
    s'&=\frac{rs}{1+r(s-1)}.\label{eq:coordinate_transformation2}
\end{align}
\end{subequations}
To demonstrate how this coordinate transformation can be used to derive identities for $R_{J,k}(r)$, consider the maximally mixed state in the symmetric subspace $\hat\rho_{0}$ and the state which is an equal mixture of the lowest weight in each SU(2)$_{\{23\}}$ irrep, $\hat\rho_\pi$. One observes that $\hat\rho_\pi=\hat U_\pi^{\{12\}}\hat\rho_0(\hat U_\pi^{\{12\}})^\dagger$, and thus these states are related by the coordinate transformation in Eqs.~(\ref{eq:coordinate_transformation1}-\ref{eq:coordinate_transformation2}). Calculating each of their Wigner functions $W_0$ and $W_\pi$ we find that
\begin{subequations}
\begin{align}
    W_0(s,r)=&\frac{1}{\sqrt{N+1}}R_{N/2,0}(r),\label{eq:w0}\\
    \nonumber W_\pi(s,r)=&\frac{1}{N+1}\sum_{J,k}\frac{(-1)^k(2k+1)(2J)!}{\sqrt{(2J-k)!(2J+k+1)!}}\\
    &\times R_{J,k}(r)P_k(1-2s).\label{eq:wpi}
\end{align}
\end{subequations}
The coordinate transformation implies that $W_\pi(s',r')=W_0(s,r)$. Since, $W_0$ is $s$-independent, we can set $s=0$ in the coordinate transformation, leaving us with $s'=0$ and $r'=1-r$. Evaluating $W_\pi$ at these transformed coordinates and equating Eqs.~(\ref{eq:w0}) and (\ref{eq:wpi}) one finds
\begin{align}
    \nonumber &R_{N/2,0}(r)\\
    &\;=\frac{1}{\sqrt{N+1}}\sum_{J,k}\frac{(-1)^k(2k+1)(2J)!}{\sqrt{(2J-k)!(2J+k+1)!}}R_{J,k}(1-r),
\end{align}
which would be nontrivial to derive if not for covariance property. These transformation rules may also allow one to transform a difficult Wigner function computation, into a simpler one.

\section{Conclusions}
\label{sec:conclusions}

The standard SU(2) Wigner function is defined on the surface of a sphere. This Wigner function applies when the system being represented exists within a single SU(2) irreducible representation. However, in systems comprised of $N$ spin-1/2 particles many physically relevant noise sources do not act collectively on the system, and thus the noise is not total angular momentum preserving. This led us to define a Wigner function which can incorporate local symmetric noise into its description. In this work, we accomplished this by representing a noisy SU(2) system using the collective state basis \cite{Chase2008}, and embedding that basis into an SU(3) irrep.

Previous generalizations of the SU(2) Wigner function to the case of multiple SU(2) irreps have been made \cite{Klimov2008,Klimov2017}, which do so by introducing an additional index to the kernel, defining Weyl symbols for different subspaces of the total Hilbert space. In this work we took a different approach, instead using the SU(3) Wigner function construction in \cite{Klimov_SUn,Martins2019}. This approach requires assuming additional group actions on the system of $N$ spins, which do not naturally exist. 

We found that, by assuming no coupling between SU(2) irreps of different $J$, which is a physically required condition, the SU(3) Wigner function takes only 3 real parameters, which can be interpreted as a polar, azimuthal, and radial component. This leads us to imagine the topology of the space as a solid ball, rather than a sphere, inspiring the name: the \emph{solid} spin Wigner function. The measure on the space is not truly that of a solid ball, but this interpretation is useful for visualizing SU(2) systems undergoing local noise.

Finally, the question of which Lie algebra is best suited to describe noise in spin ensembles has interesting avenues of future research. While SU(3) is particularly well-suited to describe atom loss (as its symmetric subspace resembles that of the space used in \cite{Zhang2018} to describe symmetric loss and gain channels) other noise channels such as optical pumping and depolarizing are represented using a collective state basis which is subtly different. While atom loss changes total and projective angular momentum by half-integer amounts, other local noise channels change angular momentum by integer amounts. Therefore, a more suitable Lie algebra for these noise sources might be one with SU(2) subgroups which differ in integer amounts, and which have eight rays as opposed to the six rays in SU(3). A possible contender might be the $\mathfrak{so}(5)$ algebra, which has eight rays and SU(2) subgroups differing by integer amounts. Representations of $\mathfrak{so}(5)$ do not have triangular weight diagrams \cite{Rowe2012}, but this could be compensated for by restricting to a subspace of an $\mathfrak{so}(5)$ representation.

\section{Acknowledgments}
This work is supported by funding from the NSF Quantum Leap Challenge Institutes program, Award No. 2016244, and the NSF Grant PHY-2116246. I would like to thank my PhD advisor Ivan H. Deutsch for his guidance throughout this research. I would also like to thank Hubert de Guise for his insightful conversations regarding phase conventions in $\mathfrak{su}(3)$, and his willingness to share resources from previous investigations into the SU(3) Wigner function.

\bibliography{refs}

\appendix

\section{Operators on Collective State Space}
\label{app:ops_on_col}

In this appendix we will define a mapping from operators in the collective state basis $\mathcal C$ to operators on a new vector space $\mathcal V$, which we will refer to as the vectorized collective state space. We will show that by representing operators and superoperators in $\mathcal C$ using $\mathcal V$ one can ignore many of the technical difficulties that occur when using the collective state basis. Further, we will show that in $\mathcal V$ local noise channels can be formally mapped to Lindblad jump operators, as has been used computationally by groups previously \cite{Zhang2018,Forbes2024}.

\subsection{The Vectorized Collective State Basis}
\label{app:issues}

We begin by recalling that density operators in $\mathcal C$ can be defined as normalized linear combinations of the collective states basis elements,
\begin{equation}
    \col{J,M,N}{J,M',N}=\frac{1}{d_J^N}\sum_{\alpha=1}^{d_J^N}\ketbra{J,M,\alpha}{J,M',\alpha},\label{eq:coll_state_appendix}
\end{equation}
where $d_J^N$ is the degeneracy of irrep $J$ with a total number of spin-1/2 particles $N$. Arbitrary permutationally invariant states can then be written as
\begin{equation}
    \hat\rho=\sum_{J,M,M'}p_{M,M'}^{(J)}\col{J,M,N}{J,M',N}.\label{eq:coll_rho}
\end{equation}
It would be convenient if one could write $\hat\rho$ in the collective state basis as a matrix with entries $p_{M,M'}^{(J)}$ where $J$ represents block diagonal subspaces, and $M$ and $M'$ represent elements within those subspaces, i.e.
\begin{equation}
    \hat\rho=\begin{pmatrix}
  \overset{J=2}{\boxed{
    \begin{matrix}
      * & * & * & * & * \\
      * & * & * & * & * \\
      * & * & * & * & * \\
      * & * & * & * & * \\
      * & * & * & * & *
    \end{matrix}
  }} & & \\
  & \overset{J=1}{\boxed{
    \begin{matrix}
      * & * & * \\
      * & * & * \\
      * & * & *
    \end{matrix}
  }} & \\
  & & \overset{J=0}{\boxed{*}}
\end{pmatrix}\label{eq:desired_matrix}
\end{equation}
where the example above is an $N=4$ collective state space, with entries in asterisks representing $p^{(J)}_{M,M'}$. However, the collective state basis is \emph{not} a normalized basis, despite being orthogonal. This is evidenced by the fact that
\begin{align}
    \nonumber\col{J,M,N}{J,M',N}&\col{J,M',N}{J,M'',N}\\
    &=\frac{1}{d_J^N}\col{J,M,N}{J,M'',N}.\label{eq:col_on_col}
\end{align}
Thus, if one were to construct a matrix of the form Eq.~(\ref{eq:desired_matrix}) for $\hat\rho$ they would encounter problems when applying $\hat\rho$ to other operators, as the matrix in Eq.~(\ref{eq:desired_matrix}) inherently assumes that the basis elements $\col{J,M,N}{J,M',N}$ are mutually normalized. The solution to this problem is to introduce the \emph{dual} basis $\{\hat E_{M,M'}^{(J,N)}\}$ to the collective states
\begin{equation}
    \hat E_{M,M'}^{(J,N)}=d_J^N\col{J,M,N}{J,M',N},
\end{equation}
such that
\begin{align}
    \nonumber\Tr&\Big[\col{J,M,N}{J,M',N}\Big(\hat E_{M'',M'''}^{(J',N)}\Big)^\dagger\Big]\\
    &\hspace{2cm}=\delta_{J,J'}\delta_{M,M''}\delta_{M',M'''},
\end{align}
and using this, define dual expansions for the state $\hat\rho$ and operators $\hat A$ such that the space can be interpreted as being expanded in an orthonormal basis.

Consider the expansion of the density matrix $\hat\rho$ and a collective operator $\hat A$ in the collective state basis and dual basis respectively
\begin{align}
    \hat\rho&=\sum_{J,M,M'}\Tr(\hat\rho \left(\hat E_{M,M'}^{(J,N)}\right)^\dagger)\col{J,M,N}{J,M',N}\label{eq:rho_in_coll}\\
    \hat A&=\sum_{J,M,M'}\Tr(\hat A\col{J,M',N}{J,M,N})\hat E^{(J,N)}_{M,M'}.
\end{align}
We observe that when $\hat A$ is applied to $\hat\rho$ we obtain
\begin{align}
    \nonumber\hat A\hat\rho=&\sum_{J,M,M',M''}\Tr(\hat A\col{J,M,N}{J,M'',N})\\
    &\times\Tr(\hat\rho \left(\hat E_{M,M'}^{(J,N)}\right)^\dagger)\col{J,M,N}{J,M',N},\label{eq:A_on_rho}
\end{align}
which is written in the collective state basis. In fact, Eq.~(\ref{eq:A_on_rho}) looks like the result one would expect if the collective states formed a normalized basis, and both $\hat\rho$ and $\hat A$ were expanded in that basis. This observation inspires us to define a new Hilbert space $\mathcal V$ spanned by $\ket{J,M}$ with mapping rule from $\mathcal C\mapsto\mathcal B(\mathcal V)$ on operators $\hat O\mapsto\hat O_{\mathcal V}$ given by
\begin{align}
    \hat\rho_{\mathcal V}&=\sum_{J,M,M'}\Tr(\hat\rho \left(\hat E_{M,M'}^{(J,N)}\right)^\dagger)\ketbra{J,M}{J,M'}\label{eq:v_map_on_rho}\\
    \hat A_{\mathcal{V}}&=\sum_{J,M,M'}\Tr(\hat A\col{J,M',N}{J,M,N})\ketbra{J,M}{J,M'}.\label{eq:v_map_on_A}
\end{align}
By mapping density operators using the overlap with the dual basis, and collective observables using the overlap with collective state basis, we ensure that the action of collective operators on the density matrix is identical in $\mathcal V$ as it was in $\mathcal C$. We call $\mathcal V$ the \emph{vectorized collective state space}, since it resembles a promotion of collective states (which are mixed) to pure states with the same angular momentum eigenvalues $J$ and $M$.

We can define the vectorized collective state space $\mathcal V$ as a direct sum of SU(2) irreps $\mathcal H_J$
\begin{equation}
    \mathcal V=\bigoplus_{J=0}^{N/2}\mathcal H_J
\end{equation}
where the direct sum increases in half-integer increments. Unlike the tensor product space of $N$ spin-1/2 particles $\mathcal H_N$, $\mathcal V$ contains no degeneracies by design. Further, while $\mathcal H_N$ has a direct sum decomposition into irreps differing in size $J$ by integer amount, $\mathcal V$ contains irreps differing by half-integer amounts. This is a convenient definition, as it allows for more arbitrary noise channels like atom loss \cite{Zhang2018}, and makes the mapping to SU(3) more straightforward.

Before continuing, we would like to point out that the mapping $\mathcal C\mapsto\mathcal B(\mathcal V)$ on operators $\hat A$ is convenient for several reasons when $\hat A$ is a collective operator. First, collective operators are identical across degenerate subspaces within a given SU(2) subspace. Thus, from Eq.~(\ref{eq:v_map_on_A}) we find that
\begin{equation}
    \mel{J,M,\alpha}{\hat A}{J,M',\alpha}=\mel{J,M}{\hat A_\mathcal V}{J,M'}.
\end{equation}
This means that the image of collective operators in $\mathcal B(\mathcal V)$ look identical to their corresponding operators in an arbitrary degenerate irrep of $\mathcal H_N$. Further, since collective operators are block diagonal across SU(2) irreps, i.e.
\begin{equation}
    \hat A=\bigoplus_{J,\alpha}\hat A_{J,\alpha}
\end{equation}
we note that for two collective operators $\hat A$ and $\hat B$
\begin{equation}
    \mel{J,M,\alpha}{\hat A\hat B}{J,M',\alpha}=\mel{J,M}{\hat A_\mathcal{V}\hat B_\mathcal{V}}{J,M'}.
\end{equation}
Thus, once operators are mapped into $\mathcal B(\mathcal V)$, any operations involving application of collective operators to density matrices can be evaluated. One should be careful to remember, however, that repeated application of a density operator in $\mathcal B(\mathcal V)$ to itself will yield incorrect results, as any operator applied to $\hat\rho_\mathcal V$ must be mapped according to Eq.~(\ref{eq:v_map_on_A}). For example, the state $\hat\rho=\col{J,M,N}{J,M,N}$ in $\mathcal C$ is not pure for $J<N/2$, and yet it maps to $\ketbra{J,M}$ in $\mathcal B(\mathcal V)$. Thus, computing the purity of $\hat\rho$ using $\hat\rho_\mathcal{V}$ cannot be done by simply computing $\Tr(\hat\rho_\mathcal{V}^2)$. Conversely, the trace of an operator $\hat\rho\in\mathcal C$ can only be evaluated in $\mathcal V$ if it is mapped according to Eq.~(\ref{eq:v_map_on_rho}). This means that any operator $\hat A_{\mathcal V}$ mapped using Eq.~(\ref{eq:v_map_on_A}) will generally not have a trace which agrees with its counterpart in $\mathcal C$, i.e. $\Tr(\hat A)\neq\Tr(\hat A_{\mathcal V})$. Despite these subtleties, the vectorized collective state space is extremely useful for simulating the evolution of permutationally symmetric states, and calculating collective observables.

\subsection{Local Noise Maps in $\mathcal V$}
It is tempting to define jump operators $\hat L_{k,q}$ which transfer elements of the collective state basis as
\begin{align}
    \nonumber\hat L_{k,q}&\col{J,M,N}{J,M',N}\hat L_{k,q}^\dagger\\
    &\propto\col{J+k,M+q,N}{J+k,M'+q,N}.
\end{align}
Practically, such approaches have been used in previous studies \cite{Zhang2018,Forbes2024,Forbes2025}, but formally not all of these operators $\hat L_{k,q}$ can exist in the space of linear operators $\mathcal B(\mathcal H_N)$ on $\mathcal H_N$. To see why, consider some jump operator $\hat B\in\mathcal B(\mathcal H_N)$ acting on an initially pure state $\ket\psi$ in the symmetric subspace of $\mathcal H_N$. The purity $\mathcal P$ of $\ket\psi$ after a jump occurs is
\begin{equation}
    \mathcal P=\Tr[\frac{\hat B\ketbra\psi\hat B^\dagger\hat B\ketbra\psi\hat B^\dagger}{\mel{\psi}{B^\dagger B}{\psi}^2}]=1,
\end{equation}
for any jump operator $\hat B$ in the span of linear operators on $\mathcal H_N$. We compare this to our desired operator $\hat L_{-1,-1}$ acting on the pure spin coherent state $\ket{J,J}$. After the jump the post-selected state is
\begin{align}
    \nonumber\hat L_{-1,-1}&\col{J,J,N}{J,J,N}\hat L_{-1,-1}^\dagger\\
    &=\col{J-1,J-1,N}{J-1,J-1,N},\label{eq:nonpure_state}
\end{align}
but this post-jump state in Eq.~(\ref{eq:nonpure_state}) has purity $\mathcal P=1/d_J^N$. Thus, $\hat L_{-1,-1}$ does not exist in $\mathcal B(\mathcal H_N)$.

This issue does not occur in $\mathcal B(\mathcal V)$, as the promotion of each collective state to a pure state means that $\hat L_{k,q}$ can easily be defined using the diads of $\mathcal B(\mathcal V)$ in the $\ket{J,M}$ basis. For example, an operator which performs the desired action in Eq.~(\ref{eq:nonpure_state}) could be constructed as
\begin{equation}
    \hat L_{k,q}=\ketbra{J-1,M-1}{J,M}.
\end{equation}
Using $\mathcal V$, collective jump operators which represent the action of local jumps can be constructed using spherical tensors, as discussed in Sec. \ref{sec:collective_states}.

\begin{widetext}
\section{Reducing the $\mathfrak{su}(3)$ Clebsch-Gordan Coefficient and Irreducible Tensor}
\label{sec:reducing_A}

From ref.~\cite{deGuise2020} we know that $\mathfrak{su}(3)$ CG coefficients can be factored into an $\mathfrak{su}(2)$ CG coefficient, multiplied by a ``reduced'' coefficient $A$ which depends only on the numbers defining rank and irrep-type. In the following we will use this to simplify the irreducible tensor from Eq.~(\ref{eq:su3_tensor}). From \cite{Martins2019} we can write that the SU(3) irreducible tensor coupling $(\lambda,0)$ to $(0,\lambda)$ to obtain $(\sigma,\sigma)$ is
\begin{equation}
    \hat T^\lambda_{\sigma;\gamma I_\gamma}=\sum_{\nu\beta}\ketbra{(\lambda,0);\nu;I_\nu}{(\lambda,0)\beta;I_\beta}\tilde C_{\lambda\nu I_\nu;\lambda^*\beta^*I_\beta}^{\sigma\gamma I_\gamma},
\end{equation}
where the star indicates the conjugate irrep and conjugate weight ($\beta_i\to \lambda-\beta_i$), and the tilde implies a phase difference between the coefficient in the above expression and a true $\mathfrak{su}(3)$ CG. App. A of \cite{Martins2019} states
\begin{equation}
    \tilde C_{\lambda\alpha I_\alpha;\lambda^*\beta^*I_\beta}^{\sigma\nu I_\nu}=(-1)^{\lambda-\beta_2}\left\langle\begin{array}{cc|c}
        (\lambda,0) & (0,\lambda) & (\sigma,\sigma)\\
        \alpha I_\alpha & \beta^* I_\beta & \nu I_\nu
\end{array}\right\rangle,
\end{equation}
where the bracketed term is an $\mathfrak{su}(3)$ CG [Eq.~(\ref{eq:su3_cg})] representing the overlap between the tensor product of $\ket{(\lambda,0)\alpha I_\alpha}$ and $\ket{(\lambda,0)\beta I_\beta}$ with $\ket{(\sigma,\sigma)\nu I_\nu}$, and $\alpha$, $\beta$, $\nu$ represent the 3-dimensional occupation number vector defined in Sec.~\ref{sec:su3}. We are further given that
\begin{align}
    \left\langle\begin{array}{cc|c}
         (\lambda,0) & (0,\lambda) & (\sigma,\sigma)  \\
         (\alpha_1\alpha_2\alpha_3)I_\alpha& (\beta_1\beta_2\beta_3)I_\beta & (c_1c_2c_3)k
    \end{array}\right\rangle&=\left\langle\begin{array}{cc||c}
         (\lambda,0)&(0,\lambda) & (\sigma,\sigma)  \\
         a_1 I_\alpha & \beta_1 I_\beta & c_1 k
    \end{array}\right\rangle\left\langle\begin{array}{cc|c}
        I_\alpha & I_\beta & k \\
        (\alpha_2-\alpha_3)/2 & (\beta_2-\beta_3)/2 & (c_2-c_3)/2 
    \end{array}\right\rangle.
\end{align}
Next, we define some of these parameters in terms of the collective state basis description, which can be done by inspection of weight diagrams such as Fig.~\ref{fig:rotation_diagram}. We write that
\begin{subequations}
\begin{align}
    \lambda&=N\\
    I_\alpha&=I_\beta=J\\
    \alpha_2&=J+M\\
    \alpha_3&=J-M\\
    \beta_2&=N-(J+M')\\
    \beta_3&=N-(J-M')
\end{align}
\end{subequations}
and using the property of occupation numbers that $\nu_1+\nu_2+\nu_3=\lambda+2\mu$ we can solve for $\alpha_1$ and $\beta_1$ as
\begin{equation}
    \alpha_1=N-2J\hspace{2cm}\beta_1=2J.
\end{equation}
Substituting in these definitions we obtain
\begin{align}
    \underset{\text{SU(3) CG coefficient}}{\left\langle\begin{array}{cc|c}
         (\lambda,0) & (0,\lambda) & (\sigma,\sigma)  \\
         (\alpha_1\alpha_2\alpha3)I_\alpha& (\beta_1\beta_2\beta_3)I_\beta & (c_1c_2c_3)k
    \end{array}\right\rangle}&=\underset{A_{N,J}^{\sigma,k}}{\left\langle\begin{array}{cc||c}
         (N,0)&(0,N) & (\sigma,\sigma)  \\
         (N-2J) J & (2J)J & \sigma k
    \end{array}\right\rangle}\underset{\text{SU(2) CG coefficient}}{\left\langle\begin{array}{cc|c}
        J & J & k \\
        M & -M' & M-M' 
    \end{array}\right\rangle}.
\end{align}
Above we have identified the reduced coefficient $A_{N,J}^{\sigma,k}$, which can be written as \cite{Martins2019}
\begin{align}
    \nonumber A_{J,N}^{\sigma,k}=&(-1)^N\frac{\sigma+1}{\sigma!}\sqrt{\frac{2(2J+1)(N-\sigma)!(\sigma-k)!(\sigma+k+1)!}{(N+\sigma+2)!}}\\
    &\times\sum_{\nu}(-1)^{\nu}\frac{\nu!(N-\nu+\sigma+1)!}{\left(\nu-(N-2J)\right)!(N-2J+\sigma-\nu)!(N-\nu)!(\nu-\sigma)!}
    \begin{Bmatrix}
        \frac12(N+\sigma-\nu)&\frac12(N-\nu)&\sigma/2\\
        \frac12(\sigma-\nu+N-2J)&\frac12(\nu-N+2J)&\sigma/2\\
        J&J&k
    \end{Bmatrix},\label{eq:A_reduced}
\end{align}
where the bracketed term is a 9j symbol and the sum runs over all allowed values of $\nu$ such that the factorial arguments are nonnegative and the 9j symbol is triangular.

Substituting these expressions into the SU(3) irreducible tensor we obtain
\begin{align}
    T^\lambda_{\sigma;\gamma I_\gamma}&=\sum_{J,M,M'}(-1)^{N-J-M'}A_{N,J}^{\sigma,k}\left\langle\begin{array}{cc|c}
        J & J & k \\
        M & -M' & M-M' 
    \end{array}\right\rangle\ketbra{J,M}{J,M'}\\
    &=(-1)^N\sum_{J,q}(-1)^{-J-q}A_{J,N}^{\sigma,k}\sum_{M}(-1)^{-M}C_{J,M;J,-(M+q)}^{k,-q}\ketbra{J,M}{J,M+q}\\
    &=(-1)^N\sum_{J,q}(-1)^{q}A_{J,N}^{\sigma,k}\sqrt{\frac{2k+1}{2J+1}}\sum_M (-1)^{2M} C_{J,M,k,q}^{J,M+q}\ketbra{J,M}{J,M+q}\\
    &=(-1)^N\sum_{J,q}(-1)^{2J+q}A_{J,N}^{\sigma,k}\left(T_{k,q}^J\right)^\dagger\\
    &=\sum_{J,q}(-1)^{N+2J+q}A_{J,N}^{\sigma,k}\left(T_{k,q}^J\right)^\dagger,
\end{align}
where $T_{k,q}^J$ is an SU(2) spherical tensor on an SU(2) irrep of size $J$. Finally, we note that the factor of $(-1)^N$ results in a kernel with negative trace for $N$-odd. Thus, without consequence to the required properties of the tensor or kernel, we define the tensor in Eq.~(\ref{eq:su3_tensor}) with this phase factor neglected.
\end{widetext}

\section{Normalization and Traciality}
\label{app:normalization}

In this appendix we want to address a subtlety of the normalization and traciality conditions of Wigner functions, and explicitly describe how these properties appear in the SU(3) Wigner function. While one might usually expect the Wigner function to be normalized such that it integrates to 1, in reality it only needs to be normalized to a constant. For example, in the formulation of the SW conditions used in this work, we treated normalization as requiring that the kernel has unit trace. By calculating the Wigner function of the maximally mixed state one finds that the integral of $W(\theta,\phi,r)$ is
\begin{equation}
    \int \Tr(\hat\omega(\theta,\phi,r)\mathds1/d) r(2\pi)\mathrm dr\mathrm d\theta\mathrm d\phi=4\pi^2/d\equiv\mathcal N
\end{equation}
where $d$ is the dimension of $(\lambda,0)$, and thus this constant $\mathcal N$ is the volume of the SU(3) Wigner function. Further, traciality is often stated as
\begin{equation}
    \int_MW_A(\Omega)W_B(\Omega)\mathrm d\Omega=\Tr(\hat A\hat B)\label{eq:traciality}
\end{equation}
as we defined it in Sec. \ref{sec:formalism}. In reality, this constant factor relating to the norm may appear in the traciality condition if the Wigner function is not normalized to 1. To see why, consider, for example, the trace of the maximally mixed state $W(\theta,\phi,r)=1/d$ with itself. Traciality implies that this should yield the purity of the maximally mixed state, $\mathcal P=1/d$. However, one finds that
\begin{equation}
    \int\frac{1}{d^2}r(2\pi)\mathrm dr\mathrm d\theta\mathrm d\phi=\frac{4\pi^2}{d^2}=\mathcal N\times\mathcal P
\end{equation}
and thus the traciality condition explicitly reads
\begin{equation}
    \int_MW_A(\Omega)W_B(\Omega)\mathrm d\Omega=\mathcal N\Tr(\hat A\hat B).
\end{equation}
In some contexts, like the bosonic Wigner function, the kernel for state operators is sometimes normalized by $\mathcal N$ while the kernel for the operators is not. This convention, while leading to two slightly different mappings for state operators and non-state operators, means that the volume of a state's Wigner function is normalized to 1, and that the traciality condition works exactly as Eq.~(\ref{eq:traciality}) if one of the Wigner functions represents a state.

\end{document}